\documentclass{elsart}
\usepackage{graphicx}
\usepackage{amsmath}
\begin{document}
\begin{frontmatter}
\title{Geometric tri-product of the spin domain and Clifford algebras}
\author{Yaakov Friedman }
  \address{Jerusalem College of
Technology,\\ P.O.B. 16031, Jerusalem 91160 Israel\\email:
friedman@jct.ac.il}

%

\begin{abstract}
We show that the triple product defined by the spin domain
(Bounded Symmetric Domain of type 4 in Cartan's classification) is
closely related to the geometric product in Clifford algebras. We
present the properties of this tri-product and compare it with the
geometric product.

 The spin domain can be used to construct a
model in which spin 1 and spin1/2 particles coexist. Using the
geometric tri-product, we develop the geometry of this domain. We
present a geometric spectral theorem for this domain and obtain
both spin 1 and spin 1/2 representations of the Lorentz group on
this domain.



\end{abstract}

\end{frontmatter}
\section{Introduction}

The spin factor, a bounded symmetric domain of type IV in the
Cartan classification \cite{CE35}, plays an important role in
physics. It was shown in \cite{FR92} that the state space of any
two-state quantum system is the dual of a complex spin factor. In
\cite{F04} and \cite{FS} it was shown that a new dynamic variable,
called $s$ velocity, which is a relativistic half of the usual
velocity, is useful for solving explicitly relativistic dynamic
equations. In \cite{F04} it was shown that automorphism group of
this $s$ velocity coincide with the conformal group and its Lie
algebra is described by the triple product defined uniquely by the
spin domain. The basic operators of the complex spin triple
product are closely related to the geometric product of Clifford
algebras.

 We start by defining the spin triple product, which we
also call the geometric tri-product, and discuss its connection to
the geometric product in Clifford algebras. Then we study the
algebraic properties and the geometry of the unit ball of the spin
factor and its dual. Here, the duality between minimal and maximal
tripotents plays a central role.  In particular, this duality
enables us to construct both spin 1 \textit{and} spin 1/2
representations of the Lorentz group on the same spin factor.
Thus, we can incorporate particles of integer and half-integer
spin in one model. As a result, the complex spin factor with its
triple product is a new model for supersymmetry.

Most of the results of this article appear in full detail in
Chapter 3 of \cite{F04}.

\section{The geometric tri-product of the spin domain}

Let $\mathbf{C}^n$ denote $n$-dimensional (finite or infinite)
complex Euclidean space with the natural basis
\[
\mathbf{e}_1=(1,0,\ldots,0),\mathbf{e}_2=(0,1,\ldots,0),\dots,\mathbf{e}_n=(0,\ldots,0,1)
\] and the usual inner product
\begin{equation}\label{innerprod} \langle\mathbf{a}|\mathbf{b}\rangle=
a_1\overline{b}_1+a_2\overline{b}_2+\cdots+a_n\overline{b}_n,\end{equation}
 where
$\mathbf{a}=(a_1,\ldots,a_n),\; \mathbf{b}=(b_1,\ldots,b_n)$.
 The Euclidean norm of $\mathbf{a}$ is defined by
 $|\mathbf{a}|=\langle\mathbf{a}|\mathbf{a}\rangle^{1/2}$.
 As in \cite{FR01}, \cite{F04},  we  define  a triple product on $\mathbf{C}^n$ by
 \begin{equation}\label{maincomplspintrip}
\{\mathbf {a},\mathbf{b},\mathbf{c}
\}=\langle\mathbf{a}|\mathbf{b} \rangle\mathbf
{c}+\langle\mathbf{c} |\mathbf{b}\rangle\mathbf{a} -\langle\mathbf
{a}|\mathbf{\overline{c}}\rangle \overline{\mathbf{b}},
\end{equation} where $\overline{\mathbf
{b}}=(\overline{b}_1,\ldots,\overline{b}_n)$ denotes the complex
conjugate of $\mathbf {b}$. In  \cite{F04} this product is called
the \textit{spin triple product}\index{spin triple product}. In
this paper we will also call it also the \textit{geometric
tri-product}\index{geometric tri-product}.

Note that the geometric tri-product is linear in the first and
third variables ($\mathbf{a}$ and $\mathbf{c}$) and conjugate
linear in the second variable ($\mathbf{b}$). Since, by the
definition of the inner product, we have $\langle\mathbf
{a}|\mathbf{\overline{c}}\rangle=\langle\mathbf
{c}|\mathbf{\overline{a}}\rangle$, the triple product is symmetric
in the outer variables, \textit{i.e.,}
\begin{equation}\label{complspinsymerty}
 \{\mathbf {a},\mathbf{b},\mathbf{c}\}=\{\mathbf {c},\mathbf{b},\mathbf{a}
\}.
\end{equation}
The space ${ C}^n$ with the geometric tri-product is called the
\index{triple factor!spin} \index{spin factor}\textit{complex spin
triple factor} and will be denoted by $\mathcal{S}^n$. We use this
name because if we define a norm based on this triple product,
then the unit ball of $\mathcal{S}^n$ is a domain of Cartan type
IV known as the spin factor.

 The \index{spin factor!real} \textit{real part of the spin factor},
denoted $\mathcal{S}^n_\mathbf{R}$, is the subspace of
$\mathcal{S}^n$ defined by
\[\mathcal{S}^n_\mathbf{R}=\{\mathbf {a} \in \mathcal{S}^n:\;\;\;
\overline{\mathbf {a}}={\mathbf {a}}\},\] or, equivalently,
\begin{equation}\label{realpartsin}
\mathcal{S}^n_\mathbf{R}= \mbox{span}_\mathbf{R} \{\mathbf{e}_j\}.
\end{equation}
This subspace is identical, as a linear space, to $R^n$ and its
tri-product  defines the Lie algebra of the \textit{conformal
group} \index{conformal group}of the unit disk in $R^n$.

For any $ \mathbf{a},\mathbf{b} \in \mathcal{S}^n$, we define a
complex linear map
$D(\mathbf{a},\mathbf{b}):\mathcal{S}^n\rightarrow \mathcal{S}^n$
by
\begin{equation}\label{Ddef}
D(\mathbf{a},\mathbf{b})\mathbf{z}=\{\mathbf{a},\mathbf{b},\mathbf{z}\}=\langle\mathbf{a}|\mathbf{b}
\rangle\mathbf {z}+\langle\mathbf{z} |\mathbf{b}\rangle\mathbf{a}
-\langle\mathbf {a}|\mathbf{\overline{z}}\rangle
\overline{\mathbf{b}}.
\end{equation}
 The linear map defined by $D(\mathbf{a},\mathbf{b})$ can be
 expressed in the language of Clifford algebras by
\begin{equation}\label{tripleGP}
 D(\mathbf{a},\mathbf{b})=\langle\mathbf{a}|\mathbf{b} \rangle \,I +\mathbf{a}\wedge\mathbf{b},
\end{equation}
where $I$ denotes the identity operator and \[
(\mathbf{a}\wedge\mathbf{b})(\mathbf{z})=\langle\mathbf{z}
|\mathbf{b}\rangle\mathbf{a} -\langle\mathbf
{a}|\mathbf{\overline{z}}\rangle \overline{\mathbf{b}}.\]

 By a
commonly used identity \cite{Hes03}, in the real case
$\mathbf{a}\wedge\mathbf{b}$ coincides with the \textit{wedge
(exterior) product}\index{wedge product} of vectors:  \[
\mathbf{z}\cdot(\mathbf{a}\wedge\mathbf{b})=(\mathbf{z}\cdot
\mathbf{a})\mathbf{b} -(\mathbf{z}\cdot\mathbf{b})\mathbf {a}.\]
Thus, the map $D(\mathbf{a},\mathbf{b})$ resembles the
\index{product!geometric} geometric product of $\mathbf{a}$ and
$\mathbf{b}$, defined by
\begin{equation}\label{gp}
  \mathbf{a}\mathbf{b}=\langle\mathbf{a}|\mathbf{b}\rangle +\mathbf{a}\wedge \mathbf{b},
\end{equation}
where the sum of a scalar $\langle\mathbf{a}|\mathbf{b}\rangle$
and the antisymmetric bivector $\mathbf{a}\wedge \mathbf{b}$
 belongs to  the Clifford
algebra. Hence, the operator $D(\mathbf{a},\mathbf{b})$ is a
natural operator on the spin factor and plays a role similar to
that of the geometric product.

\section{ The  canonical basis of $\mathcal{S}^n$.}\label{s.3.1.2}

The \textit{canonical anticommutation relations}\index{canonical
anticommutation relations} (CAR) are the basic relations used in
the description of fermion fields.

We will show now that the natural basis of $\mathcal{S}^n$
satisfies a triple analog of the CAR. Recall that the classical
definition of CAR involves a sequence $p_k$ of elements of an
associative algebra which satisfy the relations
\begin{equation}\label{car}
 p_lp_k+p_kp_l=2\delta_{kl},
\end{equation}
where $ \delta_{kl}$ denotes the Kronecker delta.
 This implies that $p_k^2=1,$ and, therefore,
\begin{equation}\label{tcreq1}
p_kp_kp_l=p_l,\quad \mbox{ for any } 1\le k,l \le n.
\end{equation}
 Multiplying (\ref{car}) on the
 left by $p_l$, we get
\begin{equation}\label{tcreq2}
p_lp_kp_l=-p_k,\quad \mbox{ for } k\ne l.
\end{equation}
 We call the relations (\ref{tcreq1}) and (\ref{tcreq2})
the \index{canonical anticommutation relations!triple}
\textit{triple canonical anticommutation relations }\- (TCAR).

Using definition (\ref{maincomplspintrip}) of the spin triple
product,
 it is easy to verify that the elements $\mathbf{e}_1,\mathbf{e}_2,\ldots,\mathbf{e}_n$
of the natural basis of the spin triple factor satisfy the
following relations:
\begin{equation}\label{tbasis1}
 \{\mathbf{e}_l,\mathbf{e}_k, \mathbf{e}_l \}=-\mathbf {e}_k,\;\;\;\mbox{ for }k\ne
 l,\end{equation}
\begin{equation}\label{tbasis2}  \{\mathbf{e}_l,\mathbf{e}_k,\mathbf{e}_k \}=
  \{\mathbf{e}_k, \mathbf{e}_k, \mathbf{e}_l \}=\mathbf{e}_l,\;\;\mbox{
for any}\;\;  k,l,\end{equation}
\begin{equation}\label{tbasis3}\{\mathbf{e}_l,\mathbf{e}_k, \mathbf{e}_m\}=0,\quad  \mbox{for }\;
k,l,m \;\mbox{distinct}.
\end{equation}
Thus, the natural basis of the spin triple factor $\mathcal{S}^n$
or $\mathcal{S}^n_\mathbf{R}$ satisfies the TCAR. Conversely, if
we define a ternary operation on
$\{\mathbf{e}_l,\mathbf{e}_2,\ldots,\mathbf{e}_n\}$ which
satisfies (\ref{tbasis1})-(\ref{tbasis3}), then the resulting
triple product on $\mathcal{S}^n$ will be exactly the spin triple
product.

We say that a basis
$\{\mathbf{u}_1,\mathbf{u}_2\ldots,\mathbf{u}_n\}$ of
$\mathcal{S}^n$ is a \index{basis!canonical}\textit{canonical
basis of} $\mathcal{S}^n$ or a \textit{TCAR basis}
\index{basis!TCAR} if it satisfies
(\ref{tbasis1})-(\ref{tbasis3}).
 It can be shown  that
any TCAR basis is an \index{basis!orthonormal} \textit{orthonormal
basis} of $\mathbf{C}^n$. The converse, however, is not true.

Let $\mathbf{u}_l,\mathbf{u}_k$ be any two distinct elements  of
an orthonormal basis in $R^n$ spanning a plane
$\Pi=\mbox{span}_{R}\{\mathbf{u}_l,\mathbf{u}_k\}$ in $R^n$. Then,
the \textit{generator of rotations}\index{rotation!generator} in
$\Pi$ is defined by
\begin{equation}\label{gen of rotation def}
 J(\mathbf{u}_l)=-\mathbf{u}_k,\;\;J(\mathbf{u}_k)=\mathbf{u}_l,
 \mbox{  and } J(\mathbf{u}_m)=0 ,\quad  \mbox{for }\;
k,l,m \;\mbox{distinct}.
\end{equation}
From (\ref{tbasis1})-(\ref{tbasis3}), it follows that the operator
$D(\mathbf{u}_l,\mathbf{u}_k)$ is the generator of rotation in the
plane $\Pi=\mbox{span}_{R}\{\mathbf{u}_l,\mathbf{u}_k\}$ of
$\mathcal{S}^n$, implying that
\begin{equation}\label{rotation with D}
 \exp(\theta D(\mathbf{u}_l,\mathbf{u}_k))\;\mbox{ = rotation in
 $\Pi$  of $\mathcal{S}^n$ by the angle }-\theta .
\end{equation}\index{rotation!operator}
Note that properties (\ref{tbasis1})-(\ref{tbasis3}) can be
derived from the requirement that the operator
$D(\mathbf{u}_l,\mathbf{u}_k)$ is the generator of rotation in the
plane $\Pi=\mbox{span}_{R}\{\mathbf{u}_l,\mathbf{u}_k\}$ of
$\mathcal{S}^n$.  This result is similar to the fact \cite{Baylis}
that bivectors play the role of generators of rotations.

 On the other hand, since
$D(\mathbf{u}_k,i\mathbf{u}_k)=-iD(\mathbf{u}_k,\mathbf{u}_k)=-i\,I$,
we have
\begin{equation}\label{rotation complex}
  \exp(\theta D(\mathbf{u}_k,i\mathbf{u}_k))\mathbf{a}=e^{-i\theta}\mathbf{a}
\end{equation}
for any $\mathbf{a}\in \mathcal{S}^n$. This shows that
$D(\mathbf{u}_k,i\mathbf{u}_k)$ is the generator of the rotation
representing the action of $U(1)$ on $\mathcal{S}^n$.

 To define \textit{reflection in a plane}\index{reflection}, we will use the tri-product operator $Q(\mathbf{u})$
 defined by
\begin{equation}\label{Q oper} Q(\mathbf{u})\mathbf{a}=\{\mathbf{u},\mathbf{a},\mathbf{u}\}\end{equation}
 for $\mathbf{a}\in \mathcal{S}^n$.
 Direct calculation using the definition (\ref{maincomplspintrip}) of the geometric
 tri-product shows that if $|\mathbf{u}|=1,$ then
\begin{equation}\label{reflection with respect to plane}
 Q(\mathbf{u})\mathbf{a}=\{\mathbf{u},\mathbf{a},\mathbf{u}\}=2\langle\mathbf{u}|\mathbf{a}\rangle\mathbf{u}-
 \mathbf{a}=2P_{\mathbf{u}}\mathbf{a}-\mathbf{a} ,
\end{equation}
where $P_{\mathbf{u}}\mathbf{a}$ denotes the orthogonal projection
of $\mathbf{a}$ onto the direction of $\mathbf{u}$. Note that for
the 3D-space $\mathcal{S}_\mathbf{R}^3$ the operator
$-Q(\mathbf{u})$ defines the space reflection with respect to the
plane with normal $\mathbf{u}$. In $\mathcal{S}^n$, $180^\circ$
rotation in the plane
$\Pi=\mbox{span}_{R}\{\mathbf{u}_k,\mathbf{u}_l\}$ is given by
\begin{equation}\label{reflecion in plane}
Q(\mathbf{u}_l)Q(\mathbf{u}_k):\;\mathbf{a}\;\rightarrow\;\{\mathbf{u}_l,\{\mathbf{u}_k,\mathbf{a},\mathbf{u}_k\},\mathbf{u}_l\}.
\end{equation}
Moreover, rotation operator in the above plane $\Pi$ by an angle
$\theta$ can be obtained as a \textrm{double reflection
}\index{rotation!double reflection}in two planes with an angle
$\theta /2$ between their normals, as
\begin{equation}\label{rotation by reflection}
 Q(\exp(\frac{\theta}{2}D(\mathbf{u}_l,\mathbf{u}_k)\mathbf{u}_k)Q(\mathbf{u}_k).
\end{equation}

The natural morphisms of the complex spin triple factor
$\mathcal{S}^n$ are \index{automorphism!${\rm\mbox{Taut}}\,
(\mathcal{S}^n)$} the linear, invertible maps (bijections)
$T:\mathcal{S}^n \rightarrow \mathcal{S}^n$ which preserve the
triple product. This means that
\begin{equation}\label{compspintripauto}
 T\{\mathbf{a},\mathbf{b},\mathbf{c}\}=\{T\mathbf{a},T\mathbf{b},T\mathbf{c}\}.
\end{equation}
Such a linear map is called a \textit{triple automorphism} of
$\mathcal{S}^n$.
 We denote by $\mbox{Taut}\, (\mathcal{S}^n )$ the
group of all triple automorphisms of $\mathcal{S}^n$.

Since the definition of a TCAR basis involves only the triple
product, it is  obvious that a triple automorphism $T$ maps a TCAR
basis into a TCAR basis. In particular, the image of the natural
basis $\{\mathbf{e}_1,\mathbf{e}_2,\ldots,\mathbf{e}_n \}$ is a
TCAR basis. It can be shown  that a bijective map $T$  of the spin
triple factor $\mathcal{S}^n$ preserves the triple product,
\textit{i.e.} $T\in \mbox{Taut}\, (\mathcal{S}^n )$, if and only
if it has the form $T=\lambda U$, where $\lambda$ is a complex
number of absolute value 1 and $U$ is orthogonal. Thus,
\begin{equation}\label{Tautdef}
\mbox{Taut}\, (\mathcal{S}^n )=U(1)\times O(n),
\end{equation}
where $U(1)$ is the group of rotations in the complex plane and
$O(n)$ is the orthogonal group of dimension $n$. Thus, ${Taut}\,
(\mathcal{S}^n)$ is a Lie group with real dimension
$\frac{n(n-1)}{2}+1$.

This group is a natural candidate for the description of the state
space of a quantum system. The state description of a quantum
system is often given by a complex-valued wave function $\psi
(\mathbf{r})$, where $\mathbf{r}\in R^3$. This description is
invariant under the choice of the orthogonal basis in $R^3$,
implying that there is a natural action of the group $O(3)$ on the
state space. In the presence of an electromagnetic field, the
gauge of the field induces a multiple of the state by a complex
number $\lambda,\; |\lambda|=1$, which will not affect any
meaningful results. Multiplication of all $\psi (\mathbf{r})$ by
such a $\lambda$ corresponds to an action of the group $U(1)$ on
this state space. Moreover, even without gauge invariance, all
meaningful quantities in quantum mechanics are invariant under
multiplication by a complex number of absolute value $1$,
resulting in an action of $U(1)$. Thus, $\mbox{Taut}\,
(\mathcal{S}^n)$ acts naturally on the state space of quantum
systems. A similar result holds for quantum fields.

\section{Tripotents and singular decomposition in $\mathcal{S}^n$}

The building blocks of binary operations,  are the projections.
These are the \textit{idempotents} \index{tripotent} of the
operation, that is, non-zero elements $p$ that satisfy $p^2=p.$
For a ternary operation, the building blocks are the
\textit{tripotents}, non-zero elements $\mathbf{u}$ satisfying
$\{\mathbf{u},\mathbf{u},\mathbf{u}\}=\mathbf{u}.$ To describe the
tripotents $\mathbf{u}\in \mathcal{S}^n,$  we introduce the notion
of determinant for elements of $\mathcal{S}^n.$ For any
$\mathbf{a} \in \mathcal{S}^n$, the \index{determinant!spin}{\it
determinant} of $\mathbf{a}$, denoted $\det\mathbf{a}$, is
\begin{equation}\label{detdef} \det\mathbf{a}
=\langle\mathbf{a}|\overline{\mathbf{a}}\rangle=\sum_{i=1}^{n}
a_i^2.
\end{equation}
In case the elements of $\mathcal{S}^n$ can be represented by
matrices, this definition agrees with the ordinary determinant of
a matrix. This definition is similar to the notion of metric on a
paravector space (see \cite{Baylis}). Note that elements with zero
determinant are called \textit{null-vectors} in the literature.

The properties of the tripotents $\mathbf{u}\in \mathcal{S}^n$ are
summarized in Table \ref{TripotentsInSpinFactorTable}:
\begin{table}[h!]
  \centering
 \scalebox{0.5}{\includegraphics{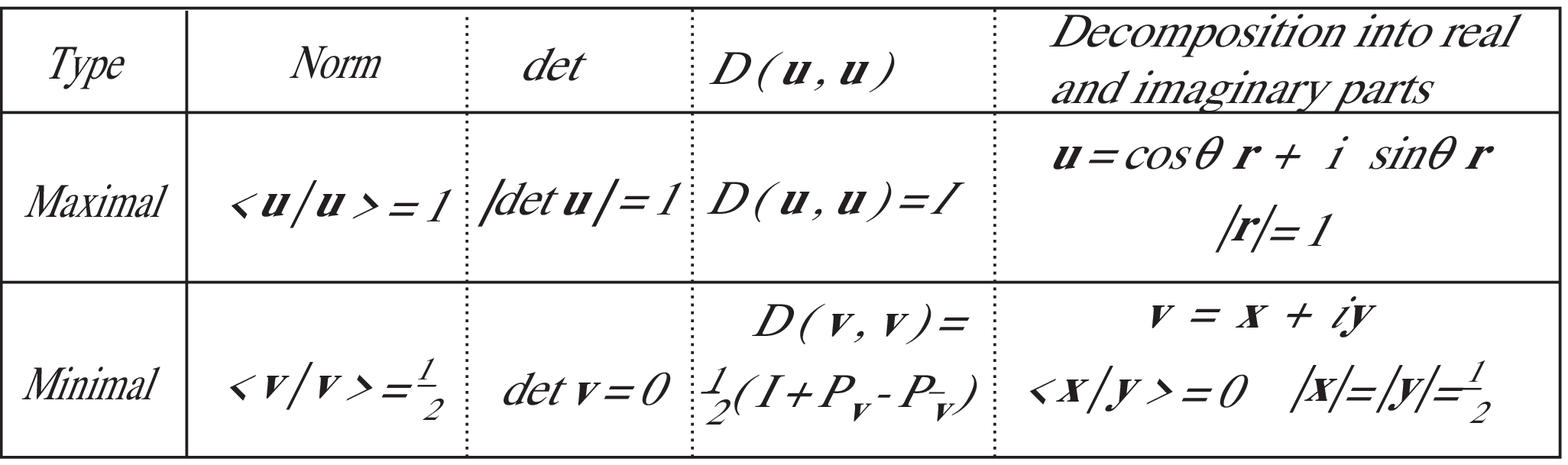}}
  \caption{The algebraic properties of tripotents in $\mathcal{S}^n.$
  }\label{TripotentsInSpinFactorTable}
\end{table}

 There are only  two types of tripotents in $\mathcal{S}^n$,{\em maximal}
 and  \textit{minimal}\index{tripotent!maximal}\index{tripotent!minimal}. The Euclidian norm of a maximal
tripotent is 1, while the norm of a minimal tripotent is
$1/\sqrt{2}$. For a maximal tripotent $\mathbf{u}$, we have $|\det
\mathbf{u}|=1$, while for a minimal tripotent $\mathbf{v}$, we
have $\det \mathbf{v}=0.$ The operator $D(\mathbf{u}, \mathbf{u})$
for a maximal tripotent $\mathbf{u}$ is the identity operator. For
a minimal tripotent $\mathbf{v}$, the operator $D(\mathbf{v},
\mathbf{v})$ is
\begin{equation}\label{dopermintripspin}
D(\mathbf{v},
\mathbf{v})=\frac{1}{2}(I+P_{\mathbf{v}}-P_{\overline{\mathbf{v}}}),
\end{equation} where $P_{\mathbf{v}}$ and
$P_{\overline{\mathbf{v}}}$ denote the orthogonal projections on
${\mathbf{v}}$ and $\overline{\mathbf{v}}$ respectively.

If we decompose a minimal tripotent $\mathbf{v}$ into real and
imaginary parts as
\begin{equation}\label{mindec}
\mathbf{v}=\mathbf{x}+i\mathbf{y}, \quad\;\; \mathbf{x},\mathbf{y}
\in \mathcal{S}^n_\mathbf{R},
\end{equation} then, $\langle\mathbf{x}
|\mathbf{y} \rangle=0$ and $|\mathbf{x}| = |\mathbf{y}|=1/2$.
Similarly, if we decompose a maximal tripotent $\mathbf{u}$ into
real and imaginary parts,  then there exist a real $\theta$ and
$\mathbf{r}\in \mathcal{S}_\mathbf{R}^n$ with $|\mathbf{r}|=1$
such that
\begin{equation}\label{maxdec}
\mathbf{u} = \cos \theta\,\mathbf{r}+i\sin \theta\,\mathbf{r}=
e^{i\theta}\mathbf{r}.
\end{equation}

The spectrum of the operator $D(\mathbf{v},\mathbf{v})$ is the set
  $\{1,1/2,0\}$, where the eigenvalue 1 is
obtained on multiples of $\mathbf{v}$ (\textit{i.e.}, on the image
of $P_{\mathbf{v}}$), the eigenvalue 0 is obtained on multiples of
$\overline{\mathbf{v}}$ (on the image of
$P_{\overline{\mathbf{v}}}$) and the eigenvalue $1/2$ is obtained
on the image of the projection
$I-P_{\mathbf{v}}-P_{\overline{\mathbf{v}}}.$ Let
$P_1(\mathbf{v})$, $P_{1/2}(\mathbf{v})$ and $P_0(\mathbf{v})$ be
the projections onto the $1$, $1/2$ and $0$ eigenspaces of
$D(\mathbf{v},\mathbf{v})$, respectively:
\begin{equation}\label{piercep1/20}
P_1(\mathbf{v})=P_{\mathbf{v}}, \;\;
P_{1/2}(\mathbf{v})=I-P_{\mathbf{v}}-P_{\overline{\mathbf{v}}},\;\;
P_0(\mathbf{v})=P_{\overline{\mathbf{v}}}.
\end{equation}
Then
\begin{equation}\label{dvaspierceproject}
 D(\mathbf{v})= P_1(\mathbf{v})+\frac{1}{2} P_{1/2}(\mathbf{v}).
\end{equation}

Since
\begin{equation}\label{piercedecproj}
 I=P_1(\mathbf{v})+P_{1/2}(\mathbf{v})+P_0(\mathbf{v}),
\end{equation}
these projections induce a decomposition of $\mathcal{S}^n$ into
the sum of the three eigenspaces:
\begin{equation}\label{piercedecomp}
 \mathcal{S}^n=\mathcal{S}^n_1(\mathbf{v})+\mathcal{S}^n_{1/2}(\mathbf{v})
 +\mathcal{S}^n_0(\mathbf{v}).
\end{equation}
This is called the \index{Peirce!decomposition}\textit{Peirce
decomposition}\index{decomposition!Peirce} of $\mathcal{S}^n$ with
respect to a minimal tripotent $\mathbf{v}.$ A very useful result
for calculations is the \textit{Peirce
calculus}\index{Peirce!calculus} formula. Let $j,k,l \in
\{1,\frac{1}{2},0\}$ with $j-k+l \in \{1,\frac{1}{2},0\}$, then
\begin{equation}\label{piercecalculusspin}
 \{ \mathcal{S}^n_j(\mathbf{v}),
\mathcal{S}^n_k(\mathbf{v}), \mathcal{S}^n_l(\mathbf{v})\} \subset
\mathcal{S}^n_{j-k+l}(\mathbf{v}).
\end{equation}
Otherwise, $\{ \mathcal{S}^n_j(\mathbf{v}),
\mathcal{S}^n_k(\mathbf{v}), \mathcal{S}^n_l(\mathbf{v})\}=0.$

We will say that two tripotents $\mathbf{v}$ and $\mathbf{u}$ are
\textit{algebraically orthogonal} \index{tripotents!algebraically
orthogonal}(denoted by $\mathbf{v}\perp\mathbf{u}$) if
\begin{equation}\label{alebr orth}
 D(\mathbf{v},\mathbf{v})\mathbf{u}=0,\mbox{  or
 }D(\mathbf{u},\mathbf{u})\mathbf{v}=0.
\end{equation}
It can be shown that $\mathbf{v}$ and $\mathbf{u}$ are
algebraically orthogonal if and only if
\begin{equation}\label{zero-divizors}
 \mathbf{v}\perp\mathbf{u}\quad \Leftrightarrow\quad D(\mathbf{v},\mathbf{u})=0.
\end{equation}
Such tripotents are the analog of nonzero zero-divisors in an
algebra. For a minimal tripotent $\mathbf{v}$, its complex adjoint
$\overline{\mathbf{v}}$ is also a minimal tripotent and is
algebraically orthogonal to $\mathbf{v}$. Moreover, and
$\mathbf{v}+\overline{\mathbf{v}}$ is a maximal tripotent. Thus,
for a minimal tripotent $\mathbf{v}$
\begin{equation}\label{min tripot orth}
 D(\mathbf{v},\overline{\mathbf{v}})=0,\;\;\mbox{ and }\;\;
  D(\mathbf{v}+\overline{\mathbf{v}},\mathbf{v}+\overline{\mathbf{v}})=I.
\end{equation}

Let $\mathbf{a}$ be any element in $\mathcal{S}^n.$ If
 $\det \mathbf{a}= 0$, then $\mathbf{a}$ is a positive multiple of
 a minimal tripotent. In fact,
 $\mathbf{u}:=\frac{1}{\sqrt{2}}\frac{\mathbf{a}}{|\mathbf{a}|}$ is a minimal
 tripotent. If $\det \mathbf{a}\ne 0$, it can be shown
 that there exist an algebraically orthogonal pair $\mathbf{v}_1,
 \mathbf{v}_2$ of minimal tripotents and a pair of non-negative real numbers
$s _1, s_2$, called the \index{singular numbers} \textit{singular
numbers} of $\mathbf{a}$, such that $s _1  \ge  s_2 \ge 0$ and
\begin{equation}\label{singular} \mathbf{a}=s _1 \mathbf{v}_1
+s_2 \mathbf{v}_2.
\end{equation} This decomposition is called the
\index{decomposition!singular}
\textit{singular decomposition of}
$\mathbf{a}.$  If $\mathbf{a}$ is not a multiple of a maximal
tripotent, then $s_1>s_2$ and the decomposition is unique. If
$\mathbf{a}$ \textit{is} a multiple of a maximal tripotent, then
$s_1=s_2$, and the decomposition is, in general, not unique.

The singular numbers satisfy
\begin{equation}\label{determinantands1s2}
  |\mbox{det}\, \mathbf{a}|=s_1s_2,
\end{equation} which corresponds to the fact that the  determinant of
a positive operator is the product of its eigenvalues, and
\begin{equation}\label{s1s2asfuncofa}
 s_1\pm s_2=\sqrt{2|\mathbf{a}|^2\pm2 |\mbox{det}\, \mathbf{a}|}.
\end{equation} Given
$\mathbf{a}$ with singular decomposition (\ref{singular}), we have
\begin{equation}\label{qubesingular}
  \mathbf{a}^{(3)}:=\{\mathbf{a},\mathbf{a},\mathbf{a}\}=s_1^3\mathbf{v}_1
+s_2^3\mathbf{v}_2,
\end{equation}
showing that the cube of an element $\mathbf{a}\in \mathcal{S}^n $
can be calculated by cubing its singular numbers. Similarly,
taking \textit{any} odd power of $\mathbf{a}$ is equivalent to
applying this odd power to its singular numbers. Moreover, for any
analytic function $f:R_+\to R_+$ we can define
\begin{equation}\label{functcalc}
  f(\mathbf{a})=f(s_1)\mathbf{v}_1
+f(s_2)\mathbf{v}_2,
\end{equation}
which play the analog of \textit{operator function}
\index{operator function} calculus.

For an arbitrary element $\mathbf{d}\in \mathcal{S}^n$, it can be
shown that  the spectrum of the linear operator
$D(\mathbf{d},\mathbf{d})$ is non-negative and that
\begin{equation}\label{mainspin}
 D(\mathbf{d},\mathbf{d})\{\mathbf{a},\mathbf{b},\mathbf{c}\}
=\{D(\mathbf{d},\mathbf{d})\mathbf{a}, \mathbf{b}, \mathbf{c}\}
-\{\mathbf{a},D(\mathbf{d},\mathbf{d}) \mathbf{b}, \mathbf{c}\} +
\{\mathbf{a}, \mathbf{b},D(\mathbf{d},\mathbf{d}) \mathbf{c}\}.
\end{equation}
This is called the \index{main identity} \textit{main identity} of
the triple product. Note that from this identity follows that
$D(\mathbf{d},\mathbf{d})$ is a \textit{triple
derivation}\index{triple!derivation} and thus its exponent is a
\textit{triple automorphism}\index{triple!automorphism}.

\section{Geometry of the unit ball of the spin factor}

Let $\mathbf{a}\in \mathcal{S}^n$ have singular decomposition
(\ref{singular}). We define a new norm, called the \index{operator
norm}\index{norm!operator!spin} \textit{operator norm of}
\textbf{a}, by
\begin{equation}\label{norm defspin}
 \|\mathbf{a}\|=s_1.
\end{equation}
This norm satisfies the triple analog of the ``star identity,"
namely
$\|\{\mathbf{a},\mathbf{a},\mathbf{a}\}\|=\|\mathbf{a}\|^3$.
Moreover,
\begin{equation}\label{normasfuncof aspin}
  \|\mathbf{a}\|= \frac{1}{\sqrt{2}}(\sqrt{|\mathbf{a}|^2+ |\det
   \mathbf{a}|}+\sqrt{|\mathbf{a}|^2-  |\det
   \mathbf{a}|}).
\end{equation}
The operator norm can also be defined by
\begin{equation}\label{normspind}
  \|\mathbf{a}\|^2=\|D(\mathbf{a},\mathbf{a})\|_{op},
\end{equation}
where $\|D(\mathbf{a},\mathbf{a})\|_{op}$ denotes the operator
norm of $D(\mathbf{a},\mathbf{a}).$ Since
\begin{equation}\label{normcompare}
|\mathbf{a}|\le \|\mathbf{a}\|\le \sqrt{2}|\mathbf{a}|,
\end{equation}
the operator  norm is equivalent to the Euclidean norm on
$\mathbf{C}^n.$

\begin{figure}[h!]
  \centering
  \scalebox{0.5}{\includegraphics{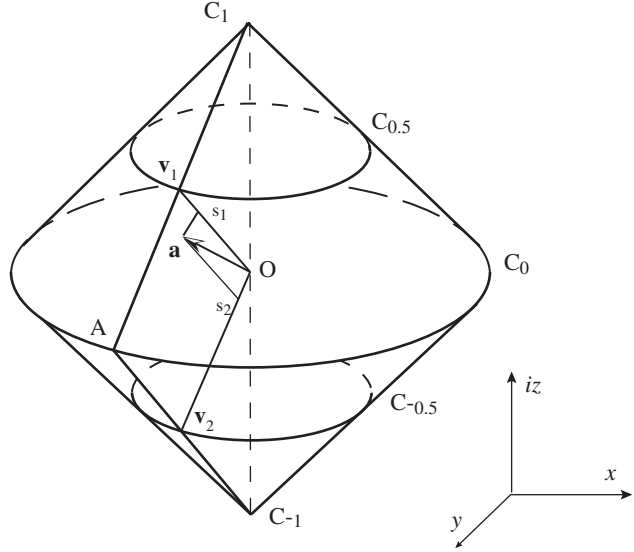}}
  \caption[The intersection of $D_{s,n}$ with the
subspace $(x,y,iz).$]{The domain $D_1$ obtained by intersecting
$D_{s,n}$ with the subspace $M_1=\{(x,y,iz):\;x,y,z\in R\}.$ $D_1$
is the intersection of two circular cones. The
\index{tripotent!minimal}minimal tripotents belong to two circles
$C_{0.5}$ and $C_{-0.5}$, whose respective equations are $iz=0.5$
and $iz=-0.5$ . The \index{tripotent!maximal}maximal tripotents
are the two points $C_1=(0,0,i)$ and $C_{-1}=(0,0,-i),$ as well as
the points of the circle $C_0$: $iz=0$ . The norm-exposed faces
are either points or line segments. }\label{3dspinoper}
\end{figure}

We denote the unit ball of  $\mathcal{S}^n$ by
\begin{equation}\label{compspinunitball}
 D_{s,n}=\{\mathbf{a} \in \mathcal{S}^n:\;\;\|\mathbf{a}\|\le 1\}.
\end{equation}
The intersection of this ball with $\mathcal{S}^n_\mathbf{R}$ is
the Euclidean unit ball  of $R^n.$  The  geometry of $D_{s,n}$ is
non-trivial. To gain an understanding of this geometry, we
consider a three-dimensional section $D_1 $ obtained by
intersecting $D_{s,n}$ with the real subspace
$M_1=\{(x,y,iz,0,...):\;x,y,z\in R\}.$ Each element of $\mathbf{a}
\in D_1$ is of the form $\mathbf{a} =(x,y,iz,0,...).$ It can be
shown  that
\begin{equation}\label{normspinsect}
 ||\mathbf{a}||=\sqrt{x^2+y^2}+|z|.
\end{equation}
Thus,\[D_1=\{(x,y,iz,0,...):\;\;\sqrt{x^2+y^2}\le 1-|z|\},\] which
is a double cone (see Figure \ref{3dspinoper}).

To locate the minimal tripotents $\mathbf{v}$ in $D_1$, we
introduce polar coordinates $r, \theta$ in the $x$-$y$ plane. Then
$\mathbf{v}=(r \cos  \theta, r \sin \theta,iz,0,...).$  The
conditions $\det\mathbf{v}=0$ and
$|Re(\mathbf{v})|=|Im(\mathbf{v})|=1/2$ imply that
\begin{equation}\label{mintripspinsect}
\mathbf{v}=1/2(\cos \theta ,\sin \theta, \pm i),
\end{equation}
so the minimal tripotents lie on  two circles $C_{0.5}$ and
$C_{-0.5}$ of radius 1/2. Maximal tripotents are multiples of a
real vector of unit length. Thus, the maximal tripotents of $D_1$
are $C_1=(0,0,i,)$, and $C_{-1}=(0,0,-i)$ and the circle
$C_0=\{(\cos \theta, \sin \theta,0):\theta \in R \}$ of radius 1.

We can now visualize the \textit{geometry of the singular
decomposition}\index{decomposition!singular!geometry of}. Let
$\mathbf{a}=(r\cos \theta ,r \sin \theta, iz)$ and let $r>z>0.$
The minimal tripotents in the singular decomposition of
$\mathbf{a}$ are
\[\mathbf{v}_1=1/2(\cos \theta , \sin \theta, i),\;\;
\mathbf{v}_2=1/2(\cos \theta , \sin \theta, -i).\] These
tripotents are the intersection of the plane through $\mathbf{a}$,
$C_{1}$ and $C_{-1}$ with the circles $C_{0.5}$ and $C_{-0.5}$ of
minimal tripotents .  The \textit{singular numbers}
\index{singular numbers} of $\mathbf{a}$ are $s_1=r+z$ and
$s_2=r-z.$  Thus the \index{decomposition!singular}singular
decomposition of $\mathbf{a}$ is
\[\mathbf{a}=\frac{r+z}{2}(\cos \theta , \sin \theta, i)
+\frac{r-z}{2}(\cos \theta , \sin \theta,-i).\] See Figure
\ref{3dspinoper}.

\section{The homogeneity of the unit ball of $\mathcal{S}^n$}

Let $D$ be a domain in a complex linear space $X$.  A map $\xi :D
\rightarrow X$  is called a \textit{vector field}\index{vector
field!analytic} on $D$. We will say that a vector field $\xi$ is
\textit{analytic} if, for any point $\mathbf{z}\in D,$ there is a
neighborhood of $\mathbf{z}$ in which $\xi$ is the sum of a power
series. A well-known theorem from the theory of differential
equations states that if $\xi$ is an analytic vector field on $D$
and $\mathbf{z}\in D,$ then the initial-value problem
\begin{equation}\label{initialvaluespin}
 \frac{d
  \mathbf{w}(\tau)}{d\tau} =\xi(\mathbf{w}(\tau)),\;\;\;\mathbf{w}(0)=\mathbf{z}
\end{equation}
has a unique solution for real $\tau$ in some neighborhood of
$\tau =0.$ A analytic vector field $\xi$ is called a
\textit{complete}\index{vector field!complete} if for any
$\mathbf{z}\in D$ the solution of the initial-value problem
(\ref{initialvaluespin}) exists for \textit{all} real $\tau$ and
$\mathbf{w}(\tau)\in D.$

For each $\mathbf{a}\in \mathcal{S}^n,$ we define a vector field
representing a generator of translation, by
\begin{equation}\label{gentranslspin}
 \xi_\mathbf{a}(\mathbf{w})=\mathbf{a}-\{\mathbf{w},\mathbf{a},\mathbf{w}\}.\;\;
 \end{equation}
 Note that $\xi_\mathbf{a}$ is a
second-degree polynomial in $\mathbf{w}$ and, thus, an analytic
vector field. Since $\xi_\mathbf{a}$ is tangent on the boundary of
$D_{s,n},$ the solution of (\ref{initialvaluespin}) exists for any
real $\tau$. This solution generates an analytic map defined  by
$\exp(\xi_\mathbf{a}\tau)(\mathbf{z})=\mathbf{w}(\tau)$ for any
$\mathbf{z}\in D_{s,n},$ where $\mathbf{w}(\tau)$ denote a
solution of (\ref{initialvaluespin}) with $\xi=\xi_\mathbf{a}$.

For any given $\mathbf{b}\in D_{s,n}$ with the singular
decomposition $\mathbf{b}=s_1\mathbf{v}_1+s_2\mathbf{v}_2$, let  $
\mathbf{a}=\tanh ^{-1}(s_1)\mathbf{v}_1+\tanh
^{-1}(s_2)\mathbf{v}_2,$  then $\exp
(\xi_\mathbf{a})(\mathbf{0})=\mathbf{b}.$ This show that the
origin in domain $D_{s,n}$ can be moved to any point
$\mathbf{b}\in D_{s,n}$ by an analytic automorphism of this
domain. This property of a domain is called\textit{ homgeneouity }
\index{domain!homgeneous} of the domain.

Since $D_{s,n}$ is the unit ball in the  operator norm, the
reflection map $\mathbf{w}\to -\mathbf{w}$ is an analytic symmetry
on $D_{s,n}$ which fixes only the origin. Clearly, $D_{s,n}$ is
bounded. Since  $D_{s,n}$ is homogeneous, it is a
\index{domain!bounded symmetric} \textit{bounded symmetric
domain}, meaning that for any $\mathbf{a}\in D_{s,n}$, there is an
analytic automorphism $s_\mathbf{a}$ which is a symmetry
(\textit{i.e.} $s_\mathbf{a}^2=id$) and fixes only the point
$\mathbf{a}$. It is known that any bounded symmetric domain define
uniquely a triple product for which the generators of translations
are given by (\ref{gentranslspin}). Thus, the spin triple product
is defined uniquely by the domain $D_{s,n}$.

For any pair of points $\mathbf{a},\mathbf{b}$ in a bounded
symmetric domain there is an operator, called the \index{Bergman
operator} \textit{Bergman operator}, defined as
\begin{equation}\label{Bergman}
 B(\mathbf{a},\mathbf{b})=
 I-2D(\mathbf{a},\mathbf{b})+Q(\mathbf{a})Q(\mathbf{b}).
\end{equation}
It can be shown \cite{L77} that for any $\mathbf{a}\in D_{s,n}$
the operator $B(\mathbf{a},\mathbf{a})$ is invertible and the
\textit{invariant metric} on $D_{s,n}$ at $\mathbf{a}$ is given
by:
\begin{equation}\label{invar metric}
  h_\mathbf{a}(\mathbf{x},\mathbf{y})=\langle
  B(\mathbf{a},\mathbf{a})^{-1}\mathbf{x}|\mathbf{y}\rangle
\end{equation}
for any $\mathbf{x},\mathbf{y}\in \mathcal{S}^n$, which can be
considered as the tangent space to $D_{s,n}$ at  $\mathbf{a}$. The
\textit{curvature tensor}\index{curvature tensor} of this metric
at 0 is given by
\begin{equation}\label{curvature}
  R_0(\mathbf{x},\mathbf{y})=
  D(\mathbf{y},\mathbf{x})-D(\mathbf{x},\mathbf{y}).
\end{equation}

\section{Geometry of the dual ball of the spin factor}

Every normed linear space $A$ over the complex numbers equipped
with a norm has a \textit{dual space},\index{dual space} denoted
$A^*,$ consisting of complex linear functionals, \textit{i.e.,}
linear maps from $A$ to the complex numbers. We define a norm on
$A^*$ by
\begin{equation}\label{dualnprmdef}
 \|f\|_*= \sup \{|f(\mathbf{w})|:\;\mathbf{w}\in A,
 \|\mathbf{w}\|\le 1\}.
\end{equation}
The dual  (or predual) of $\mathcal{S}^n$ is the set of complex
linear functionals on $\mathcal{S}^n$. We denote it by
$\mathcal{S}^n_*.$

 We use the inner  product on $\mathbf{C}^n$ to define an
imbedding of $\mathcal{S}^n$ into $\mathcal{S}^n_*,$ as follows.
For any element $\mathbf{a}\in \mathcal{S}^n,$ we define a complex
linear functional $\hat{\mathbf{a}} \in \mathcal{S}^n_*$ by
\begin{equation}\label{functionalonspindef}
 \hat{\mathbf{a}}
 (\mathbf{w})=\langle\mathbf{w}|2\mathbf{a}\rangle.
\end{equation}
 The coefficient 2 of
$\mathbf{a}$ is needed to make the dual of a minimal tripotent
have norm 1. Conversely, for any $\mathbf{f} \in \mathcal{S}^n_*,$
there is an element $\check{\mathbf{f}}\in \mathcal{S}^n$ such
that for all $\mathbf{w} \in \mathcal{S}^n,$
\begin{equation}\label{checkspindef}
 \mathbf{f}(\mathbf{w})=\langle\mathbf{w}|2\check{\mathbf{f}}\rangle.
\end{equation}

The above (\ref{dualnprmdef}) norm on $\mathcal{S}^n_*$ is called
the \textit{trace norm} \index{trace norm} and is as follows. Let
$\mathbf{f}\in \mathcal{S}^n_*$. Suppose that $\check{\mathbf{f}}$
has the singular decomposition (\ref{singular})
$\check{\mathbf{f}}=s_1\mathbf{v_1}+s_2\mathbf{v}_2$.  It can
then, be shown that
\begin{equation}\label{statenormspin}
 \|\mathbf{f}\|_*=s_1+s_2=\sqrt{2|\check{\mathbf{f}}|^2+2|\det
 \check{\mathbf{f}}|}.
\end{equation}
From this, it follows that if $\mathbf{f}\in \mathcal{S}^n_*$ has
trace norm one, then
\begin{equation}\label{pure state decomp}
  \mathbf{f}=s_1\hat{\mathbf{v}}_1+s_2\hat{\mathbf{v}}_2,\;\;
  s_1,s_2\geq 0,\;s_1+s_2=1,
\end{equation}
meaning that any norm one state is a convex combination of two
algebraically orthogonal extreme states, which correspond to pure
states in quantum mechanics.

\begin{figure}[h!]
  \centering
  \scalebox{0.4}{\includegraphics{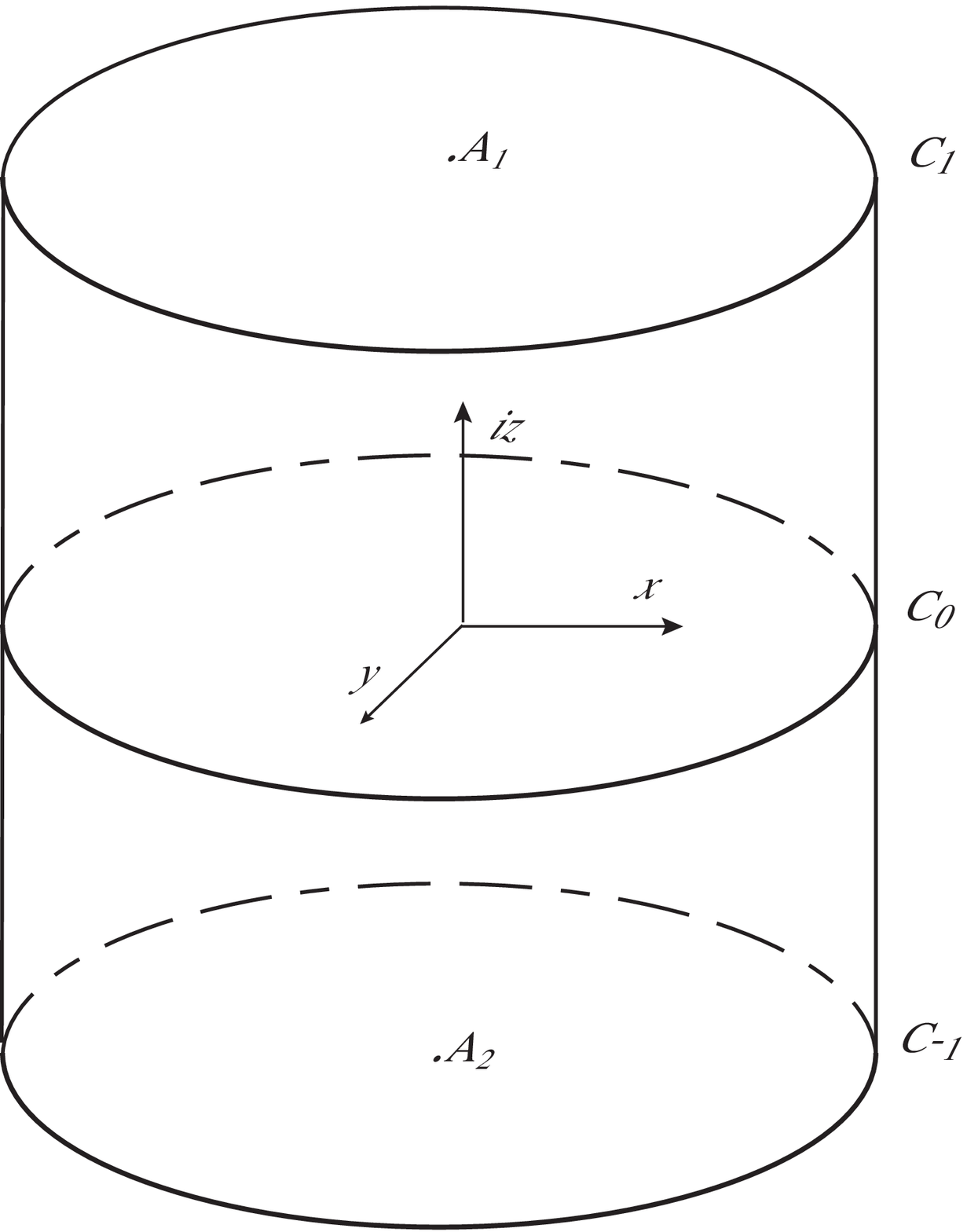}}
  \caption[The intersection of the
  state space $S_n$ with the
subspace $(x,y,iz).$]{The domain $D^*_1$ obtained by intersecting
the
  state space $S_n$ with the
subspace $M_1=\{(x,y,iz):\;\;\;x,y,z\in R\}.$   $D^*_1$ is a
cylinder. The \index{pure state}pure states, corresponding to
minimal tripotents, are extreme points of the domain and
\index{tripotent!minimal} belong to two unit circles $C_{1}:iz=1$
and $C_{-1}:iz=-1$. The functionals corresponding to
\index{tripotent!maximal}maximal tripotents are $A_1=(0,0,i)$ and
$A_{2}=(0,0,-i)$ and each point of the circle $C_0: iz=0.$  They
are centers of faces. The norm-exposed faces are either points,
line segments or disks. }\label{3dspinstate}
\end{figure}

The unit ball $S_n$ in $\mathcal{S}^n_*$ is defined by
\begin{equation}\label{statespace}
 S_n=\{\mathbf{f} \in \mathcal{S}^n_*:\; \|\mathbf{f} \|_*\le 1
 \}.
\end{equation}
We call the ball $S_n$  the \index{state space} \textit{state
space} of $\mathcal{S}^n.$   To understand the geometry of this
ball, we will examine the three-dimensional section $D^*_1$
consisting of those elements $\mathbf{f} \in S_n$ satisfying
\begin{equation}\label{threedimsubspace2}
2\check{\mathbf{f}} \in M_1=\{(x,y,iz):\;\;\;x,y,z\in R\}.
\end{equation}
It can be shown that
\[D^*_1=\{(x,y,iz):\;\;\max\{\sqrt{x^2+y^2},\,|z|\}\le
1 \},\] which is a cylinder (see Figure \ref{3dspinstate}).

To describe the functionals $\mathbf{f}$ in $D^*_1$ which
correspond to minimal tripotents $\mathbf{v}=\check{\mathbf{f}},$
we introduce polar coordinates $r,\theta$ in the $x$-$y$ plane.
For such functionals,  by (\ref{mintripspinsect}) we have
\begin{equation}\label{smintripspinsect}
2\check{\mathbf{f}}=(\cos \theta ,\sin \theta, \pm i),
\end{equation}
yielding two circles $C_{1}$ and $C_{-1}$ of radius 1. The
functionals corresponding to multiples of maximal tripotents are
multiples of a real vector of unit length. Thus, the norm one
functionals corresponding to multiples of maximal tripotents are
$A_1=(0,0,i),$  $A_{2}=(0,0,-i),$ which are the centers of the
two-dimensional discs of $\partial S_n$ and the center of the
circle $C_0=(\cos \theta, \sin \theta,0)$ of radius 1. See Figure
\ref{3dspinstate}.

\section{The state space of two-state systems}

We will now apply the results of the previous section  to
represent the states of quantum mechanic systems.  We will assume
that the state space $S$ is a unit ball of a Banach space, which
we will denote by $X_*$. We will consider only the geometry of the
state space that is implied by the measuring process for quantum
systems.  Recall that the state space consists of two types of
points. The first type represents \textit{mixed
states}\index{state!mixed}, which can be considered as a mixture
of other states with certain probabilities. The second type
represents \textit{pure states}\index{state!pure}, those states
which cannot be decomposed as a mixture of other states. By
definition, a pure state is an extreme point of the state space
$S.$

A physical quantity which can be measured by an  experiment is
called an \textit{observable}\index{observable}. The observables
can be represented as linear functionals $\mathbf{a}\in X$ on the
state space. This representation is obtained by assigning to each
state $\mathbf{f}\in S$ the expected value of the physical
quantity $\mathbf{a}$ when the system is in state $\mathbf{f}.$ A
measurement causes the quantum system to move into an eigenstate
of the observable that is being measured. Thus, the
\textit{measuring process}\index{measuring process} defines, for
any set $\Delta$ of possible values of the observable
$\mathbf{a}$, a projection $P_{\mathbf{a}}(\Delta)$ on the state
space, called a \index{projection!filtering}\textit{filtering
projection}. The projection $P_{\mathbf{a}}(\Delta)$ represents a
filtering device, called also a filter, that will move any state
$\mathbf{f}$ to the state $P_{\mathbf{a}}(\Delta)\mathbf{f},$ for
which the value of $\mathbf{a}$ is definitely in the set $\Delta$
and the probability of passing this filter is
$\|P_{\mathbf{a}}(\Delta)\mathbf{f}\|/\|\mathbf{f}\|.$

 Since, applying the filter a
second time will not affect the output state after  the first
application of the filter, $P_{\mathbf{a}}(\Delta)$ is a
projection. It is assumed that if the value of $\mathbf{a}$ on the
state $\mathbf{f}$ was definitely in $\Delta,$ then the filtering
projection does not change the state $\mathbf{f}.$ Such a
projection is called \textit{neutral}\index{projection!neutral}.

A {\em norm-exposed face} \index{face!norm-exposed}of the unit
ball $S$ of $X_*$ is a non-empty subset of $S$ of the form
\begin{equation}\label{norexposed face def}
 F_\mathbf{x}=\{\mathbf{f}\in S:\mathbf{f}(\mathbf{x})=1\},
\end{equation}
 where $\mathbf{x}\in X,\|\mathbf{x}\|=1$.  Recall that a {\em face}
  $G$ of a convex set $K$ is a non-empty convex subset of $K$ such that if
  $\mathbf{g}\in G$ and $\mathbf{h},\mathbf{k}\in
K$ satisfy $\mathbf{g}=\lambda \mathbf{h}+(1-\lambda)\mathbf{k}$
for some $\lambda\in (0,1)$, then $\mathbf{h},\mathbf{k}\in G$. We
say that two states $\mathbf{f},\mathbf{g}$ are orthogonal if
$\|\mathbf{f}+\mathbf{g}\|=\|\mathbf{f}-\mathbf{g}\|=\|\mathbf{f}\|+\|\mathbf{g}\|.$
For any subset $A$, $A^\Diamond$ denotes the set of all elements
which are orthogonal to every element of $A.$ An element
$\mathbf{u}\in X$ is called a {\em projective unit} if
$\|\mathbf{u}\|=1$ and $\mathbf{u}(F_{\mathbf{u}}^{\Diamond})=0$.

Motivated by the measuring processes in quantum mechanics,  we
define a {\em symmetric face}\index{face!symmetric} to be a
norm-exposed face $F$ in $S$ with the following property: there is
a linear isometry $S_F$ of $X_*$ onto $X_*$ which is a symmetry,
\textit{i.e.}, $S_{F}^2=I$, such that $\|S_F\|=1$ and the fixed
point set of $S_F$ is $(\overline{\mbox{sp}}F)\oplus
F^{\Diamond}$. The map $S_F$ is called the facial symmetry
associated with $F$. A complex normed space $X_*$ is said to be
{\em weakly facially symmetric}\index{facially symmetric!weakly}
if every norm-exposed face in $S$ is symmetric. A weakly facially
symmetric space $X_*$ is also {\em strongly facially
symmetric}\index{facially symmetric!strongly} if for every
norm-exposed face $F$ in $S$ and every $\mathbf{y}\in X$ with
$\|\mathbf{y}\|=1$ and $F\subset F_\mathbf{y}$, we have
$S_F^*\mathbf{y}=\mathbf{y}$.

 For each symmetric face $F$, we define contractive
projections $P_k(F),\ k=0,1/2,1$, on $X_*$ as follows. First,
$P_{1/2}(F)=(I-S_F)/2$ is the projection on the $-1$ eigenspace of
$S_F$. Next, we define $P_1(F)$ and $P_0(F)$ as the projections of
$X_*$ onto $\overline{\mbox{sp}}F$ and $F^{\Diamond}$,
respectively, so that $P_1(F)+P_0(F)=(I+S_F)/2$. A normed space
$X_*$ is called {\em neutral} if for every symmetric face $F$, the
projection $P_1(F)$ is neutral, meaning that
\[\|P_1(F)\mathbf{f} \|=\|\mathbf{f}\|\;\Rightarrow\;P_1(F)\mathbf{f} =\mathbf{f} \]
for any $\mathbf{f}\in X_*.$ In such spaces there is a one-to-one
correspondence between projective units and norm-exposed faces.

  In a neutral strongly
facially symmetric space $X_*$, for any non-zero element
$\mathbf{f} \in X_*$, there exists a unique projective unit
$\mathbf{v}=\mathbf{v}(\mathbf{f}),$ called the \textit{support
tripotent}\index{tripotent!support}, such that
$\mathbf{f}(\mathbf{v})=\|\mathbf{f}\|$ and $ \mathbf{v}(
\{\mathbf{f}\}^{\Diamond})=0.$ The support tripotent
$\mathbf{v}(\mathbf{f})$ is a minimal projective unit if and only
if $\mathbf{f}/\|\mathbf{f}\|$ is an extreme point of the unit
ball of $X_*$.

Let $\mathbf{f}$ and $\mathbf{g}$ be extreme points of the unit
ball of a neutral strongly facially symmetric space $X_*$. The
{\em transition probability} \index{transition probability}of
$\mathbf{f}$ and $\mathbf{g}$ is the number
 \[
 \langle \mathbf{f}|\mathbf{g}\rangle:=\mathbf{f}(\mathbf{v}(\mathbf{g})).
 \]
 A neutral strongly facially symmetric space $X_*$ is said to satisfy {\em ``symmetry of transition
 probabilities''} (STP) if for every pair of extreme points
 $\mathbf{f},\mathbf{g}\in S$, we have
 \[
 \overline{\langle \mathbf{f}|\mathbf{g}\rangle}=\langle
 \mathbf{g}|\mathbf{f}\rangle,
 \]
 where in the case of complex scalars, the bar denotes
 conjugation. We define the rank of a strongly facially
 symmetric space $X_*$ to be the maximal number of orthogonal projective units.

The two-state quantum systems are systems on which any measurement
can not give more than two different results. The state space of
such systems is of rank 2.  In \cite{FR92} it was shown that if
$X_*$ is a rank 2 neutral strongly facially symmetric space
satisfying STP, then $X_*$ is linearly isometric to the predual of
a spin factor.

\section{Spin grid of $\mathcal{S}^4$  and Pauli matrices}

Let $\{\mathbf{u}_0,\mathbf{u}_1,\mathbf{u}_2,\mathbf{u}_3\}$ be
an arbitrary TCAR basis in $\mathcal{S}^4.$ Then from
(\ref{mindec}) it follows that
\begin{equation}\label{vvtilde1}
 \mathbf{v}=0.5(\mathbf{u}_0+i\mathbf{u}_1)=0.5(1,i,0,0),\;
 \;\overline{\mathbf{v}}=0.5(\mathbf{u}_0-i\mathbf{u}_1)=0.5(1,-i,0,0)
\end{equation}
are a pair of algebraically orthogonal minimal tripotents. Note
that $\mathcal{S}^4_{1/2}(\mathbf{v})$ from the Peirce
decomposition (\ref{piercedecomp}) with respect to $\mathbf{v}$
has dimension 2. Also,
\begin{equation}\label{vvtilde2}
 \mathbf{w}=0.5(\mathbf{u}_2+i\mathbf{u}_3)=0.5(0,0,1,i),\;\;
 \overline{\mathbf{w}}=0.5(\mathbf{u}_2-i\mathbf{u}_3)=0.5(0,0,1,-i)
\end{equation}
are a pair of algebraically orthogonal minimal tripotents in
$\mathcal{S}^4_{1/2}(\mathbf{v})$.

The set
$\{\mathbf{v},\overline{\mathbf{v}},\mathbf{w},\overline{\mathbf{w}}\}$
is a basis of $\mathcal{S}^4$ consisting of minimal tripotents. It
is an example of a spin grid, which we will define now. We say
that a collection of tripotents is \textit{compatible
}\index{tripotents!compatible} if the collection of all Pierce
projections associated with this family commute. We say that a
collection of four elements
$(\mathbf{v},\overline{\mathbf{v}};\mathbf{w},\overline{\mathbf{w}})$
(composed of two pairs) in $\mathcal{S}^4$  form a \index{spin
grid}\textit{spin grid } if the following relations hold:
\begin{itemize}\label{odd quadrangle}
  \item[$\bullet$]
  $\mathbf{v},\mathbf{w},\overline{\mathbf{v}},\overline{\mathbf{w}}$ are
  minimal compatible tripotents,
  \item[$\bullet$] the pairs
  $(\mathbf{v},\overline{\mathbf{v}})$ and
  $(\mathbf{w},\overline{\mathbf{w}})$ are algebraically
  orthogonal,
  \item[$\bullet$] the pairs
  $(\mathbf{v},\mathbf{w}),(\mathbf{v},\overline{\mathbf{w}}),
  (\mathbf{w},\overline{\mathbf{v}})$ and
  $(\overline{\mathbf{w}},\overline{\mathbf{v}})$ are \textit{co-orthogonal}
  (the pair $(\mathbf{v},\mathbf{w})$ is said to be co-orthogonal \index{tripotents!co-orthogonal}if
   $D(\mathbf{v},\mathbf{v})\mathbf{w}=0.5\mathbf{w}$ and
  $D(\mathbf{w},\mathbf{w})\mathbf{v}=0.5\mathbf{v}$),
  \item[$\bullet$]
  $\{\mathbf{w},\mathbf{v},\overline{\mathbf{w}}\}=-0.5\overline{\mathbf{v}}$
  and
  $\{\mathbf{v},\mathbf{w},\overline{\mathbf{v}}\}=-0.5\overline{\mathbf{w}}.$
\end{itemize}

 The  spin grid in $\mathcal{S}^4$ is also called an \textit{odd quadrangle}\index{quadrangle!odd}.

  The following $2\times 2$ elementary matrices
\begin{equation}\label{martixrepoddquad}
 \mathbf{v}=\begin{pmatrix}
   1 & 0 \\
   0 & 0 \
 \end{pmatrix},\overline{\mathbf{v}}=\begin{pmatrix}
   0 & 0 \\
   0 & 1 \
 \end{pmatrix},\mathbf{w}=\begin{pmatrix}
   0 &1 \\
   0 & 0 \
 \end{pmatrix},\overline{\mathbf{w}}=\begin{pmatrix}
   0 & 0 \\
  -1 & 0 \
 \end{pmatrix},
\end{equation}
with the triple product
\begin{equation}\label{classicaltripleprod}
\{a,b,c\}=\frac{ab^*c+cb^*a}{2}
\end{equation}
are an example of an odd quadrangle. They are isomorphic to the
spin grid in $\mathcal{S}^4$. Their complex span is isomorphic to
the space of $2\times 2$ complex matrices, which now can be used
to represent the triple product in $\mathcal{S}^4$. The minus sign
in $\overline{\mathbf{w}}$ is needed in order that the quadrangle
will be an odd one.

Since $\mathbf{u}_0=\mathbf{v}+\overline{\mathbf{v}}$,
$\mathbf{u}_1=i(\overline{\mathbf{v}}-\mathbf{v})$,
$\mathbf{u}_2=\mathbf{w}+\overline{\mathbf{w}}$ and
$\mathbf{u}_3=i(\overline{\mathbf{w}}-\mathbf{w})$, the TCAR basis
of $\mathcal{S}^4$
$\{\mathbf{u}_1,\mathbf{u}_2,\mathbf{u}_3,\mathbf{u}_4\}$ in this
representation becomes
\[ \mathbf{u}_0=\begin{pmatrix}
   1 & 0 \\
   0 & 1 \
 \end{pmatrix},\mathbf{u}_1=\begin{pmatrix}
   -i & 0 \\
   0 & i \
 \end{pmatrix},\mathbf{u}_2=\begin{pmatrix}
   0 &1 \\
   -1 & 0 \
 \end{pmatrix},\mathbf{u}_3=\begin{pmatrix}
   0 & -i \\
  -i & 0 \
 \end{pmatrix}.\]
 Note that
\begin{equation}\label{Pauli matr repr}
 \mathbf{u}_0=I,\mathbf{u}_1=-i\sigma _3, \mathbf{u}_2=-i\sigma _2,\mathbf{u}_3=-i\sigma
_1,
\end{equation}
 where $\sigma_j$ denote the \textit{Pauli matrices}\index{Pauli matrices}. Conversely, the
elements of the spin grid
$\mathbf{v}=0.5(\mathbf{u}_0+i\mathbf{u}_1)$ and
$\overline{\mathbf{v}}=0.5(\mathbf{u}_0-i\mathbf{u}_1)$ are
obtained from the TCAR basis by formulas similar to the
\textit{creation and annihilation
operators}\index{operator!creation}\index{operator!annihilation}
in quantum field theory.

Any element $\mathbf{a}=(a_0,a_1,a_2,a_3) \in \mathcal{S}^4$ can
be represented by a $2\times 2$ matrix $A$ as
\[\sum^{4}_{j=1}a_j\,\mathbf{u}_j=\begin{pmatrix}
  a_0-ia_1 & a_2-ia_3 \\
  -a_2-ia_3& a_0+ia_1
\end{pmatrix}=A.\]
Note that
\begin{equation}\label{detmatrixspin}
  \det A=a_0^2+a_1^2+a_2^2+a_3^2=\det \mathbf{a},
\end{equation}
providing another justification for the definition of the
\index{determinant!spin}determinant in the
 spin factor.

\begin{figure}[h!]
  \centering
 \scalebox{0.45}{\includegraphics{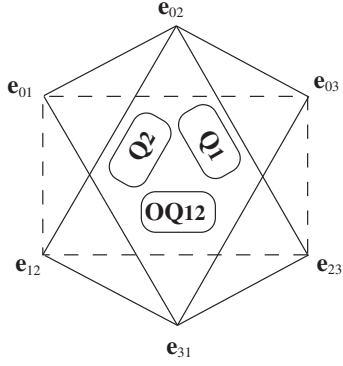}}
  \caption{Two quadrangles $\mathbf{Q}1$ and $\mathbf{Q}2$ glued by the diagonal. Here,
   $(\mathbf{e}_{01},\mathbf{e}_{23};\mathbf{e}_{03},\mathbf{e}_{12})$ form
   an odd quadrangle, denoted $\mathbf{OQ}12$.}\label{quad2}
\end{figure}

 For a spin factor $\mathcal{S}^n$ of any dimension, the spin
 grid basis is constructed from odd quadrangles and each pair of quadrangles
 is glued by the diagonal. We will demonstrate this on the spin
 factor $\mathcal{S}^6$, which will play an important role
 later. The grid of $\mathcal{S}^6$ can be represented by 6
 (coming in 3 pairs)
 elementary $4\times 4$ antisymmetric matrices:
 \[\{\mathbf{e}_{01},\mathbf{e}_{23};\mathbf{e}_{02},
 \mathbf{e}_{31};\mathbf{e}_{03},\mathbf{e}_{12}\},\]
where $\mathbf{e}_{kl}$ could be identified with
$D(\mathbf{u}_k,\mathbf{u}_l)$ as an operator on $\mathcal{S}^4$.
The spin grid consists of 3 odd quadrangles :
$(\mathbf{e}_{01},\mathbf{e}_{23};\mathbf{e}_{02},
 \mathbf{e}_{31}),$ $(\mathbf{e}_{02},
 \mathbf{e}_{31};\mathbf{e}_{03},\mathbf{e}_{12})$ and
 $(\mathbf{e}_{01},\mathbf{e}_{23};\mathbf{e}_{03},\mathbf{e}_{12})$
 glued by the diagonal (a common pair), as seen on Figure \ref{quad2}. Note that we had to
 use $\mathbf{e}_{31}=-\mathbf{e}_{13}$ to make all the
 quadrangles odd.

\section{The spin 1 Lorentz group representation on $\mathcal{S}^4$}

\index{Lorentz group!spin 1 representation}It is known that the
transformation of the electromagnetic field strength
$\mathbf{E},\mathbf{B}$ from one inertial system to another
preserves the complex quantity $\mathbf{F}^2,$ where
$\mathbf{F}=\mathbf{E}+ic\mathbf{B}.$  If we consider $\mathbf{F}$
as an element of $\mathcal{S}^3,$ then $\mathbf{F}^2=\det
\mathbf{F}.$ Thus, if we take $\mathcal{S}^3$ as the space
representing the set of all possible electromagnetic field
strengths, then the Lorentz group acts on $\mathcal{S}^3$ by
linear transformations which preserve the determinant. This leads
us to study the Lie group of determinant-preserving linear maps on
$\mathcal{S}^3$ (and, in general, on $\mathcal{S}^n$) and the Lie
algebra of this group.

Let  $\mbox{Dinv}\, (\mathcal{S}^n)$ denote the group of all
invertible linear maps  $\mathcal{S}^n \rightarrow \mathcal{S}^n$
which preserve the determinant and let $\mbox{dinv}\,
(\mathcal{S}^n)$ denote  the Lie algebra of $\mbox{Dinv}\,
(\mathcal{S}^n)$.  It can be shown that $\mbox{dinv}\,
(\mathcal{S}^n)\subset A_n({C}),$ where $A_n({C})$ denotes the
space of all $n \times n$ complex antisymmetric matrices. Using
the triple product on $\mathcal{S}^n$, we can express this Lie
algebra \index{Lie algebra!$\mbox{dinv}\, (\mathcal{S}^n)$}as
\begin{equation}\label{dinvspin}
\mbox{dinv}\, (\mathcal{S}^n)=\{\sum_{k<l}
d_{kl}D(\mathbf{u}_k,\mathbf{u}_l) :\;\: d_{kl} \in C \}.
\end{equation}

We define now a spin 1 representation of the Lorentz group by
elements of $\mbox{Dinv}\, (\mathcal{S}^4).$ Let
 $J_1,J_2,J_3,K_1,K_2,K_3$ denote the standard
infinitesimal generators of rotations and boosts respectively, in
the Lorentz Lie algebra. Furthermore, let
$\mathbf{u}_0,\mathbf{u}_1,\mathbf{u}_2,\mathbf{u}_3$ denote a
TCAR basis of $\mathcal{S}^4$. We will define a representation
$\pi_s^4$ of the Lorentz Lie algebra by elements of $\mbox{dinv}\,
(\mathcal{S}^4).$  For ease of notation, we shall write $D_{jk}$
instead of $D(\mathbf{u}_j,\mathbf{u}_k),$ for
$j,k\in\{0,1,2,3\},$ $j\ne k.$ We have seen earlier that
$D_{23}=D(\mathbf{u}_2,\mathbf{u}_3)$  generates a rotation around
the $x$-axis. Thus, we define $\pi_s^4$ by
\begin{equation}\label{pi4lorrepJ}
\pi_s^4 (J_1)=D_{23},\;\pi_s^4(J_2)=D_{31},\;\pi_s^4 (J_3)=D_{12}.
\end{equation}
Since $K_1$, the generator of a boost in the $x$-direction,
perform both an $x$-coordinate space change of the moving frame
and also a time change due to the relative speed between the
frames, it is natural to define
\begin{equation}\label{pi4lorrepK}
 \pi_s^4(K_1)=iD_{01},\; \pi_s^4(K_2)=iD_{02}, \;\pi_s^4
(K_3)=iD_{03}.
\end{equation}
The coefficient $i$ is needed to satisfy the commutation
relations.

It is easy to show that the commutation relations of the Lorentz
algebra are satisfied. The representation of the Lorentz group by
elements of $\mbox{Dinv}\, (\mathcal{S}^4)$ is now obtained by
taking the exponent of the basic elements of $\pi_s^4 .$

It is easy to check that the subspace
\begin{equation}\label{Minkowskiins4}
  M_1=\{(x^0,x^1,x^2,x^3)=x^\nu \mathbf{u}_\nu \in \mathcal{S}^4:\;\;x^0\in
  R,\;\;x^1,x^2,x^3\in iR\}
\end{equation}
is invariant under $\pi_s^4 .$  We can attach the following
meaning to the subspace $M_1.$  Let $M$ be the Minkowski space
representing the space-time coordinates $(t,x,y,z)$ of an event in
an inertial system. Define a map $\Psi: M\to M_1$ by
\begin{equation}\label{embedingspacetimes4s}
 \Psi(t,x,y,z)=
ct\mathbf{u}_0-ix\mathbf{u}_1-iy\mathbf{u}_2-iz\mathbf{u}_3.
\end{equation}
We use the minus sign for the space coordinates in order that the
resulting Lorentz transformations will have their usual form. Note
that \[\det (\Psi(t,x,y,z))=(ct)^2-x^2-y^2-z^2=s^2,\] where $s$ is
the space-time interval.

Any map $T\in \mbox{Dinv}\, (\mathcal{S}^4)$ which maps $M_1$ into
itself generates an interval-preserving map
\begin{equation}\label{lamdanadpis4}
 \Lambda=\Psi^{-1}T\Psi
\end{equation}
from $M$ to $M$. Thus, any map $T$ from $\pi_s^4$ generates by
(\ref{lamdanadpis4}) a space-time Lorentz transformation. For
example, consider the Lorentz transformation generated by a boost
$K_1$ in the $x$-direction. Obviously, such transformation changes
the $x$-coordinate and the time and do not change
$y,z$-coordinates. Direct calculation show that if $T=\exp
\varphi\pi_s^4(K_1)$, then
\[
\Lambda(t,x,y,z)=\left(\begin{array}{cccc}
\cosh \varphi&c^{-1}\sinh \varphi&0&0\\
c\sinh \varphi&\cosh \varphi&0&0\\
0&0&1&0\\
0&0&0&1
\end{array}\right)\left(\begin{array}{c}
t\\
x\\
y\\
z
\end{array}\right),
\]
which is the usual \index{Lorentz group!space-time
transformation}\textit{Lorentz space-time transformation} for the
boost in the $x$-direction, where $\tanh\varphi=\mathbf{v}/c,$ and
$\mathbf{v}$ is the relative velocity between the systems.
Conversely, any space-time Lorentz transformation $\Lambda$
generates a transformation $T=\Psi\Lambda\Psi ^{-1}$ on $M_1$
which can be extended linearly to a map on $\mathcal{S}^4$ which
belongs to $\pi_s^4.$ Thus, the usual Lorentz space-time
transformation is equivalent to a representation $\pi_s^4,$  and
$\pi_s^4$ can be considered as an extension of the usual
representation of the Lorentz group from space-time to
$\mathcal{S}^4.$

 In addition to the subspace $M_1,$ the representation $\pi_s^4$
 preserves the subspace
\begin{equation}\label{Minkowskiins42}
  M_2=\{(p_0,p_1,p_2,p_3)=p_\nu \mathbf{u}_\nu \in \mathcal{S}^4:\;\;p_0\in
  iR,\;\;p_1,p_2,p_3\in R\},
\end{equation} which is complementary to $M_{1}.$
 We can attach the following meaning to the subspace $M_2.$  Let $\widetilde{M}$
be the Minkovski space representing the four-vector momentum
$(p_0,p_1,p_2,p_3)=m_0(c\gamma,\gamma\mathbf{v}),$ where $m_0$ is
the rest-mass and $(\gamma,\gamma\mathbf{v}/c)$ is the
four-velocity of the object. Define a map $\widetilde{\Psi}:
\widetilde{M}\to M_2$ by
\begin{equation}\label{embedingspacetimes4a}
 \widetilde{\Psi}(p_0,p_1,p_2,p_3)=i\,p_0
\mathbf{u}_0+p_1\mathbf{u}_1+p_2\mathbf{u}_2+p_3\mathbf{u}_3.
\end{equation}
Note that \[\det
(\widetilde{\Psi}(p_0,p_1,p_2,p_3))=-p_0^2+p_1^2+p_2^2+p_3^2=-(E/c)^2+\mathbf{p}^2=-(m_0c)^2\]
is invariant under the Lorentz transformations.

Any linear map $T$ on $\mathcal{S}^4$ which maps  $M_2$ into
itself generates a map
\begin{equation}\label{lamdanadpis4a}
 T|_{\widetilde{M}}=\widetilde{\Psi}^{-1}T\widetilde{\Psi}
\end{equation}
from $\widetilde{M}$ to $\widetilde{M}$. If $T\in \mbox{Dinv}\,
(\mathcal{S}^4)$, then the map $T|_{\widetilde{M}}$
 is a \index{Lorentz group!four-vector momentum transformation}
 \textit{Lorentz transformation on the four-vector momentum}
space.  It can be shown that any Lorentz four-vector momentum
transformation is equivalent to an element of the representation
$\pi_s^4,$ and $\pi_s^4$ can be considered as an extension of the
usual representation of the Lorentz group from
 four-vector momentum space to $\mathcal{S}^4.$

The electromagnetic field $\bf{E,B}$ on $\widetilde{M}$ is
represented by\[\mathbf{F}=
E_k\pi_s^4(K_k)|_{\widetilde{M}}+cB_k\pi_s^4(J_k)|_{\widetilde{M}}=\]
\begin{equation}\label{elctromageticfieldrepispi4}
 \left(\begin{array}{cccc}
    0 & E_{1} & E_{2} & E_{3} \\
   -E _{1} & 0 & cB_{3} & -cB_{2} \\
    -E_{2} & -cB_{3} & 0 & cB_{1} \\
    -E_{3} &cB _{2} &-cB _{1} &0 \
  \end{array}\right),
\end{equation}
which is representation of the field by the
\textit{electro-magnetic field tensor}.\index{electro-magnetic
field tensor} \index{electromagnetic field
representation!electro-magnetic tensor}This is very natural, since
the electric field generates boosts and the magnetic field
generates rotations.

 In classical
mechanics, we use the  phase space, consisting of position and
momentum, to describe the state of a system. The properties of the
representation $\pi _s^4$ suggest that $\mathcal{S}^4$ can serve
as a relativistic analog of the phase space by representing
space-time and four-momentum on it. Note that any relativistically
invariant multiple of four-momentum can be used instead of
four-momentum. For instance, we may use four-velocity instead of
four-momentum. In order to allow transformations under which the
subspaces $M_1$ and $M_2$ are \textit{not} invariant, we have to
multiply the four-momentum by a universal constant that will make
the units of $M_1$  equal to the units of $M_2.$

We propose two models for $\mathcal{S}^4$ as the
\textit{relativistic phase space}:\index{relativistic phase space}

1) The space-momentum model for the relativistic phase space:
\begin{equation}\label{space-momentum1}
 \Omega(t,x,y,z,E/c,p_1,p_2,p_3)=\end{equation}\[(ct+\varsigma iE/c)\mathbf{u}_0+(\varsigma p_1-ix)\mathbf{u}_1+
 (\varsigma p_2-iy)\mathbf{u}_2+(\varsigma p_3-iz)\mathbf{u}_3,\]
where the universal constant $\varsigma$ transforms momentum into
length.

2) The space-velocity model for the relativistic phase space:
\begin{equation}\label{space-momentum2}
 \widetilde{\Omega}(t,x,y,z,\gamma,\gamma\mathbf{v}_1/c,\gamma\mathbf{v}_2/c,\gamma\mathbf{v}_3/c)=
 \end{equation}\[(ct+i\varpi\gamma)\mathbf{u}_0+(\varpi\gamma\mathbf{v}_1/c-ix)\mathbf{u}_1+
 (\varpi\gamma\mathbf{v}_2/c-iy)\mathbf{u}_2+(\varpi\gamma\mathbf{v}_3/c-iz)\mathbf{u}_3,\]
where the universal constant $\varpi$ transforms velocity into
length. A similar relativistic phase space, called ``velocity
phase spacetime" was used also in \cite{Schuller}.

For example, the physical meaning of a multiple of the tripotent
$\mathbf{u}=0.5(\mathbf{u}_0\pm i\mathbf{u}_1)$ by a complex
constant $\lambda$ in the space-momentum model satisfies:
\begin{equation}\label{inerprmintrip}
 \lambda\mathbf{u}=0.5\lambda(\mathbf{u}_0\pm i\mathbf{u}_1)\Rightarrow m_0=0,\;
 x=\mp ct,\;p_1=\mp E/c,\; y=z=p_2=p_3=0.
\end{equation}
and thus represents a particle with rest-mass $m_0=0$ moving with
speed $c$ in the $x$-direction. Energy and momentum are expressed
in terms of their wavelength. The $\lambda\mathbf{u}$ correspond
also \textit{directed plane waves}\index{directed plane waves},
see \cite{Baylis} (6.15).

\section{The spin $\frac{1}{2}$ Lorentz group representations on $\mathcal{S}^4$}

\index{Lorentz group!spin $\frac{1}{2}$ representation}From
(\ref{dinvspin}), it follows that  $\mbox{dinv}\, (\mathcal{S}^4)$
coincides with the space $A_4(C)$ of $4\times 4$ antisymmetric
matrices. As mentioned above, $A_4(C)$ with the triple product
(\ref{classicaltripleprod}) is isomorphic to $\mathcal{S}^6$, the
spin factor of dimension 6. The representation $\pi_s^4,$
constructed in the previous section, uses minimal tripotents
$\pi_s^4(J_k)$ and $\pi_s^4(K_k)$ of $\mbox{dinv}\,
(\mathcal{S}^4)\approx \mathcal{S}^6 $ to represent the generators
of rotations and boosts of the Lorentz group. These elements of
$\mathcal{S}^6$ form a spin grid basis. The Lorentz group
representations $\pi^+$ and $\pi^-,$ defined below, use
\textit{maximal} tripotents which form a TCAR basis of
$\mathcal{S}^6.$

Since for any $k,$ the generator of rotation $J_k$ commutes with
the generator of a boost $K_k$ in the same direction, these
operators must be represented by commuting elements in
$\mathcal{S}^6.$ There are two possibilities for such commuting
elements. The first one is to take algebraically orthogonal
elements, as in the representation $\pi_s^4,$ where $\pi_s^4(J_k)$
and $\pi_s^4(K_k)$ are algebraically orthogonal minimal tripotents
in $\mathcal{S}^6.$ But there is another possibility -
representing these generators by (complex) linearly dependent
elements of $\mathcal{S}^6.$ More precisely, we will assume that
$\pi^+(K_k)=i\pi^+(J_k)$ and $\pi^-(K_k)=i\pi^-(J_k),$ for
$k=1,2,3.$ Under these assumptions, if the representation of the
rotations satisfy the commutation relations for the generators of
rotations, all other commutation relations of the Lorentz algebra
will be satisfied automatically.

We define the representation $\pi^+$ by
\begin{equation}\label{pi plus rep}
 \pi^+(J_1)=\frac{1}{2}(D_{01}+D_{23}),\;\pi^+(J_2)=\frac{1}{2}(D_{02}+D_{31}),\;
\pi^+(J_3)=\frac{1}{2}(D_{03}+D_{12}).
\end{equation}
The definition for the representation $\pi^-$ is similar:
\begin{equation}\label{pi minus rep}
 \pi^-(J_1)=\frac{1}{2}(D_{23}-D_{01}),\;\pi^-(J_2)=\frac{1}{2}(D_{31}-D_{02}),\;
\pi^-(J_3)=\frac{1}{2}(D_{12}-D_{03}).
\end{equation}
The constant $\frac{1}{2}$ is necessary in order to satisfy the
commutation relations. For the representation $\pi ^+$ the
electromagnetic field is expressed by use of
the\index{electromagnetic field representation!Farday}
\textit{Faraday} $\mathbf{F}=\mathbf{E}+ic\mathbf{B}$  as
\begin{equation}\label{elctromageticfieldrepispi4a}
 \mathbf{F}= iF_k\pi^+(J_k)=\frac{i}{2}\left(\begin{array}{cccc}
    0 & F_{1} & F_{2} & F_{3} \\
   -F _{1} & 0 & F_{3} & -F_{2} \\
    -F_{2} & -F_{3} & 0 &F_{1} \\
    -F_{3} &F_{2} &-F _{1} &0 \
  \end{array}\right),
\end{equation}
and similarly for the representation $\pi ^-$.

Direct calculation shows that the multiplication table for
$\pi^+(J_k)$ is as in Table \ref{jkmultip}.
\begin{table}[h!]
  \centering
\begin{tabular}{|c|c|c|c|c|}\hline
      & $ 2\pi^+(J_1)$ & $2\pi^+(J_2)$ &$2\pi^+(J_3)$ \\
  \hline
  $2\pi^+(J_1) $&  $-I$ & $-2\pi^+(J_3)$ & $2\pi^+(J_2)$ \\
  $2\pi^+(J_2)$ &  $2\pi^+(J_3)$ & $-I$ & $-2\pi^+(J_1)$ \\
 $ 2\pi^+(J_3)$ & $ -2\pi^+(J_2)$& $2\pi^+(J_1)$ & $-I$ \\\hline
\end{tabular}
\caption{Multiplication table for $ \pi^+(J_k)$}\label{jkmultip}
\end{table}

The elements $\{2\pi^+(J_1),2\pi^+(J_2),2\pi^+(J_3)\}$ and
$\{2\pi^-(J_1),2\pi^-(J_2),2\pi^-(J_3)\}$ satisfy TCAR as elements
of the spin factor  $\mathcal{S}^6.$ Moreover,
\begin{equation}\label{jkforpiTCARbasis}
 \{2\pi^+(J_1),2\pi^+(J_2),2\pi^+(J_3),2\pi^-(K_1),2\pi^-(K_2),2\pi^-(K_3)\}
\end{equation}
is a TCAR basis  of $\mbox{dinv}\,
(\mathcal{S}^4)=A_4(C)=\mathcal{S}^6.$ By direct verification, we
can show that  the commutant of $\{\pi^+(J_k):k=1,2,3\}$ is
$\mbox{sp}_{C} [ \{\pi^-(J_k):k=1,2,3\}\cup\{I\}],$ which, when
restricted to real scalars, is a four-dimensional associative
algebra isomorphic to the \index{quaternions}
\textit{quaternions}. This can be seen by examining the
multiplication table \ref{jkmultip}. The commutant of
$\{\pi^-(J_k):k=1,2,3\}$ is
$\mbox{sp}_{C}[\{\pi^+(J_k):k=1,2,3\}\cup\{I\}],$ which is also,
after restriction to real scalars, isomorphic to the quaternions.
Thus, the two representations $\pi^+$ and $\pi^-$ commute.

 The above construction of the representations $\pi^+$ and $\pi^-$ from the
 representation $\pi_s^4$ can also be done via the {\it Hodge
operator},\index{Hodge operator} also called the star operator.
The representation $\pi^+$ is then the skew-adjoint part of
$\pi_s^4$ with respect to the Hodge operator, and the
representation $\pi^-$ is the self-adjoint part of $\pi_s^4$ with
respect to the Hodge operator.

Notice also that any operator of
$\{\pi^+(J_k),\pi^-(J_k):k=1,2,3\}$  has two distinct eigenvalues,
namely $\pm \frac{1}{2},$ implying that these representations are
spin $\frac{1}{2}$ representations. This is also confirmed by
direct calculation of the exponent of the generators of rotations.
For example, in the TCAR basis
$\{\mathbf{u}_0,\mathbf{u}_1,\mathbf{u}_2,\mathbf{u}_3\}$ we have
\[
R_3(\varphi)=\exp (\varphi\pi^+(J_3))=\left(\begin{array}{cccc}
\cos \frac{\varphi}{2}&0&0&\sin \frac{\varphi}{2}\\
0&\cos \frac{\varphi}{2}&\sin \frac{\varphi}{2}&0\\
0&-\sin \frac{\varphi}{2}&\cos \frac{\varphi}{2}&0\\
-\sin \frac{\varphi}{2}&0&0&\cos \frac{\varphi}{2}
\end{array}\right).
\]
This shows that the angle of rotation in the representation is
half of the actual angle of rotation.

It can be shown that the two subspaces
\[\Upsilon_1=\mbox{sp}_C\{\mathbf{v}_1,\mathbf{v}_2\}\;\; \mbox { and
}\;\;\Upsilon_2=\mbox{sp}_C\{\mathbf{v}_{-1},\mathbf{v}_{-2}\},\]
where
\[\mathbf{v}_{\pm 1}=0.5(\mathbf{u}_0\pm i\mathbf{u}_3),\;\;
\mathbf{v}_{\pm 2}=0.5(\mathbf{u}_2\pm i\mathbf{u}_1),\] are
invariant under the representation $\pi^+.$  Note that
$\mathbf{v}_1,\mathbf{v}_2,\mathbf{v}_{-1},\mathbf{v}_{-2}$ form a
basis in $\mathcal{S}^4$ consisting of minimal tripotents and form
an odd quadrangle. The representation $\pi^{+}$ leaves  both
two-dimensional complex subspaces $\Upsilon_1$ and $\Upsilon_2$
invariant, and thus we obtain two two-dimen\-sional
representations of the  Lorentz group. These representations are
related to the Pauli matrices as follows:
 \begin{equation}\label{matrix1s}
\pi^+(J_1)|_{\Upsilon_1}=i\sigma_1,\;\;\pi^+(J_2)|_{\Upsilon_1}=i\sigma_2,
\;\;\pi^+(J_3)|_{\Upsilon_1} =i\sigma_3,
\end{equation}
and \begin{equation}\label{matrix2s}
 \pi^+(J_1)|_{\Upsilon_2}=-i\sigma_1,\;\;\pi^+(J_2)|_{\Upsilon_2}=i\sigma_2,\;\;\pi^+(J_3)|_{\Upsilon_2}
=-i\sigma_3.
\end{equation}

Hence, $\pi^+$ defines the usual spin $\frac{1}{2}$ representation
on the subspace $\Upsilon_1$ via the Pauli matrices. This means
that $\xi _1 \mathbf{v}_1 +\xi_2 \mathbf{v}_2$ forms a
\textit{spinor}\index{spinor}. On the subspace $\Upsilon_2,$ the
representation $\pi^+$ acts by complex conjugation on the usual
spin $\frac{1}{2}$ representation. Hence $\eta _1 \mathbf{v}_{-1}
+\eta_2 \mathbf{v}_{-2}$ forms a\index{spinor!dotted}
\textit{dotted spinor}. This is similar to the action of the
Lorentz group on Dirac bispinors, \index{Dirac bispinors} and so
the basis
$\{\mathbf{v}_1,\mathbf{v}_2,\mathbf{v}_{-1},\mathbf{v}_{-2}\}$ of
$\mathcal{S}^4$ can serve as a basis for bispinors. Note that the
TCAR basis
$\{\mathbf{u}_0,\mathbf{u}_1,\mathbf{u}_2,\mathbf{u}_3\}$, on the
other hand, is a basis for four-vectors.

The two subspaces
\[\widetilde{\Upsilon}_1=\mbox{sp}_C\{\mathbf{v}_{-2},\mathbf{v}_1\}\;\; \mbox { and
}\;\;\widetilde{\Upsilon}_2=\mbox{sp}_C\{\mathbf{v}_{-1},\mathbf{v}_2\}\]
are invariant under the representation $\pi^-.$  Note that
$\widetilde{\Upsilon}_1,\widetilde{\Upsilon}_2$ are obtained from
the same spin grid that was used for defining the invariant
subspaces $\Upsilon_1,\Upsilon_2$ of the representation $\pi^+.$
In both cases, the invariant subspaces are obtained by
partitioning the set of four elements of the spin grid into two
pairs of non-orthogonal tripotents. Both possible partitions are
realized in the representations $\pi^+$ and $\pi^-.$
 The restriction of $\pi^-$ to the invariant
subspaces $\widetilde{\Upsilon}_1$ and $\widetilde{\Upsilon}_2$
leads to the same Pauli spin matrices as in (\ref{matrix1s}) and
(\ref{matrix2s}). Thus, the representation $\pi^-$ is a direct sum
of two complex conjugate copies of the spin $\frac{1}{2}$
two-dimensional representation given by the Pauli spin matrices.
Hence, $\pi^-$ is also a representation of the Lorentz group on
the Dirac bispinors.

We now lift the representations $\pi^+$ and $\pi^-$ from actions
on $\mathcal{S}^4$ to an action on $\mbox{dinv}\,
(\mathcal{S}^4)=\mathcal{S}^6.$ Use the TCAR the basis given by
(\ref{jkforpiTCARbasis}) of $\mathcal{S}^6$.
 Fix an action $\Lambda$ on $\mathcal{S}^4$. For any linear operator
 $T$ on $\mathcal{S}^4,$ we define a
 transformation $\Phi (\Lambda)$
 by
\[\Phi (\Lambda)T=\Lambda T \Lambda^{-1}.\] From the definition of $\mbox{Dinv}\, (\mathcal{S}^4),$ it follows that if  $\Lambda, T\in \mbox{Dinv}\,(\mathcal{S}^4),$ then also $\Phi (\Lambda)T\in \mbox{Dinv}\,(\mathcal{S}^4).$

We define the action of the rotations of $\pi^+$ on $\mbox{dinv}\,
(\mathcal{S}^4)$ by  $R_k
(\varphi)=\Phi(\mbox{exp}(\varphi\pi^+(J_k))).$   Similarly, we
define  the action of the boosts  of $\pi^+$ on $\mbox{dinv}\,
(\mathcal{S}^4)$ by  $B_k
(\varphi)=\Phi(\mbox{exp}(\varphi\pi^+(K_k))).$  With respect to
the  basis (\ref{jkforpiTCARbasis}) of
$\mbox{dinv}\,(\mathcal{S}^4),$ for $k=1$ we get
\[
{R}_1(\varphi)= \left(\begin{array}{cccc} 1&0&0&0\\ 0&\cos \varphi&\sin \varphi&0\\
0&-\sin \varphi&\cos \varphi&0\\ 0&0&0&I_3
\end{array}\right),\]\[
{B}_1(\varphi)= \left(\begin{array}{cccc} 1&0&0&0\\ 0&\cosh
\varphi&i\sinh \varphi&0\\ 0&-i\sinh \varphi&\cosh \varphi&0\\
0&0&0&I_3
\end{array}\right).
\]
This coincides with the spin 1 representation on the spin factor
$\mathcal{S}^3$, which is the complex span of
$\{\pi^+(J_1),\pi^+(J_2),\pi^+(J_3)\}\in \mbox{dinv}\,
(\mathcal{S}^4),$ and could be identified with the space of all
Faraday vectors.

\section{Discussion}

In this paper we presented the spin domain and the triple product,
called the geometric tri-product, defined by this domain. We have
seen the properties of this product and its connection to geometry
and different concepts in physics. To understand the connection of
this model to the Clifford algebras, Table \ref{comparison} below
lists various mathematical objects and contrasts how they manifest
themselves in spin factors on one hand (with reference to this
paper) and in Clifford algebras (with reference to \cite{Baylis})
on the other hand.
\begin{table}[h!]
  \centering
  \small
\begin{tabular}{|p{1.5in}|p{1.8in}|p{1.5in}|}\hline
      Object & Spin factor & Clifford algebra \\
  \hline
  1. Natural basis &  TCAR, (\ref{tbasis1}) & CAR, (1.23) \\

 2. Generator of rotation & $D(\mathbf{u}_l,\mathbf{u}_k)$, (\ref{rotation with D}) &
   Bivector $\mathbf{e}_l\mathbf{e}_k$, (1.54) \\
 3. Reflection in plane & $\mathbf{a}\to
   -Q(\mathbf{u})\mathbf{a}$, (\ref{reflection with respect to plane}) &$\mathbf{r}\to
   \mathbf{e}_l\mathbf{e}_k \mathbf{r}\mathbf{e}_l\mathbf{e}_k$,
   (1.46)\\
 4. $180^\circ$ rotation in plane & $\mathbf{a}\to
   Q(\mathbf{u}_l)Q(\mathbf{u}_k)\mathbf{a}$, (\ref{reflecion in plane}) &$\mathbf{r}\to
   \mathbf{e}_l\mathbf{e}_k \mathbf{r}\mathbf{e}_k\mathbf{e}_l$,
   (1.47)\\
 5. Rotation in plane & $\exp(\theta D(\mathbf{u}_k,\mathbf{u}_l))$, (\ref{rotation with D}) &
    $\exp (\mathbf{e}_l\mathbf{e}_k \theta)$, (1.54) \\
 6. Rotation by reflection & $Q(\exp(\frac{\theta}{2}D(\mathbf{u}_l,\mathbf{u}_k)\mathbf{u}_k)Q(\mathbf{u}_k)$
  (\ref{rotation by reflection}) &
    $\mathbf{v}\to (\mathbf{e}_{\theta/2}\mathbf{e}_3)(\mathbf{e}_3\mathbf{e}_1)\mathbf{v}$
    $(\mathbf{e}_1\mathbf{e}_3)(\mathbf{e}_3\mathbf{e}_{\theta/2})$(1.59) \\
   7. Automorphisms &$U(1)\times O(n)$ (\ref{Tautdef})& ? \\
 8. Metric & $\det \mathbf{a}$, (\ref{detdef}) & $p\overline{p}$, (1.78) \\
 9. Idempotents & 2 type tripotents, Table 1& ?\\
 & Minimal tripotents (\ref{min tripot orth}) &Projectors (1.85)\\
 & Maximal tripotents (\ref{maxdec})& Unimodular, $p\overline{p}=1$ \\
 10.  Zero-divisors & Alg. orth.
 (\ref{zero-divizors}),(\ref{min tripot orth}) & Complementary proj. (6.24),(6.26)\\
 11. Space decomp. & Peirce decomposition (\ref{piercedecomp}) &?  (6.48)\\
 12. Element decomp. &  Singular decomp. (\ref{singular}) &
  Spectral decomp. (1.86) \\
  13. Pauli matrices & (\ref{Pauli matr repr}) & (1.27)\\
  14. Generators  & FIG 3 & Fig. 2.4\\  \hline
\end{tabular}
\caption{Comparison of Spin factor and Clifford
algebra}\label{comparison}
\end{table}

It is worth comparing the representations of the geometric product
as the product in the Clifford algebra and as $D$ operators on the
complex spin triple product. In the first case, in order to
represent $n$ canonical anticommutation relations, we need an
algebra of dimension $2^n$, while in the second case, it is enough
to consider the space $\mathcal{S}^n$ of real dimension $2n,$
along with the operators defined by the spin triple product on it.

It is not obvious that there is a physical interpretation of a
bivector as an oriented area. Is there a physical meaning for a
sum of a vector and an oriented area? We can interpret a bivector
as a generator of space rotation (see \cite{Baylis}). The action
of the electromagnetic field on the ball of relativistically
admissible velocities in \cite{F04} is described by a polynomial
of degree 2, where the constant and quadratic  terms (as in
(\ref{gentranslspin})) generate a boost (caused by the electric
field), and the linear term generates a rotation (caused by the
magnetic field). For this action, the sum of a vector and a
bivector has a physical interpretation. Such sums occur also in
the descriptions of the Lie algebras of the projective and of the
conformal groups of the unit ball in $\mathbf{R}^n$. But is there
a physical interpretation for a trivector? the sum of a vector and
a trivector?

The Lorentz group is represented  by  a spin one representation
$\pi _s^4,$ defined by (\ref{pi4lorrepK}) and (\ref{pi4lorrepJ})
as operators on  $\mathcal{S}^4$. The paravector space can be
identified with the complexified subspace $M_1$ defined in
(\ref{Minkowskiins4}) by defining $1=\mathbf{u}_0$ and
$\mathbf{e}_1=i\mathbf{u}_1,\;\mathbf{e}_2=i\mathbf{u}_2,\;\mathbf{e}_3=i\mathbf{u}_3.$
The Clifford algebra $Cl_3$ represented by paravectors,  form a
complex four dimensional space, like $\mathcal{S}^4$. A spin one
representation of the Lorentz group on $Cl_3$, called the
spinorial form, is defined by transformations  (2.32) of
\cite{Baylis} $p\to LpL^\dagger$ with $L=\exp [\mathbf{W}/2].$
This representations plays a central role in Clifford algebra
applications. At this point, we do not have for the spin factor an
analog of the representation in the spinorial form. The reason for
this is that this transformation involves two different
multiplications, from the right and from the left. For the
operators on the spinors, we do not have such different
multiplications.

On the other hand, we also obtained spin-half representations $\pi
^+$ and $\pi ^-$ of the Lorentz group on the spin factor
$\mathcal{S}^4$. The same type of representation for the
eigenspinors in $Cl_3$ is obtained in (4.14) of \cite{Baylis},
based on spinorial form. The connection between these
representations is not yet obvious.

The spin triple factor arises naturally in physics. The real spin
triple product can be constructed directly from the conformal
group which represent the transformation of $s$-velocity between
two inertial systems. The complex spin triple product was used
effectively in \cite{FS} to describe the relativistic evolution of
a charged particle in mutually perpendicular electric and magnetic
fields. In the complex case, the spin triple product is built
solely on the geometry of a Cartan domain of type IV which
represents two-state systems in quantum mechanics, as was shown in
\cite{FR92}. The spin factor is a part of a larger category of
bounded symmetric domains. It appear that for a full model for
quantum mechanics, which involve state spaces of rank larger than
two, there will be a need for bounded symmetric domains of other
types. Like any bounded symmetric domain, the spin factor possess
a well-developed harmonic analysis, has an explicitly defined
invariant measure, and supports a spectral theorem as well as
quantization and representation as operators on a Hilbert space.

Since the spin factor representation is more compact, in general,
we are currently missing several techniques that play an important
role in the Clifford algebra approach. But we believe that it is
possible to overcome these difficulties. On the other hand,
 the Clifford algebra approach currently has an advantage in its
ability to express different equations of physics in a more
compact form.

\small \vskip 1pc {\obeylines \noindent Yaakov Friedman
 \noindent Department of Applied Mathematics
  \noindent Jerusalem College of Technology
   \noindent P.O.Box 16031, Jerusalem 91160, Israel
\noindent E-mail: friedman@mail.jct.ac.il
 \vskip 1pc
\vskip 6pt \noindent Submitted: September 6, 2005;


\end{document}